\documentclass[12pt]{iopart}
\usepackage{amssymb}
\usepackage{graphicx}
\usepackage{verbatim}
\usepackage{graphicx}
\usepackage{verbatim,hyperref}
\usepackage{epsfig}
\newtheorem{thm}{Statement}
\begin{document}

\title[Impossibility of thermodynamic equilibrium establishment in Universe]{Impossibility of strict thermodynamic equilibrium establishment in the accelerated Universe}

\author{Yurii Ignat'ev}

\address{Lobachevsky Institute of Mathematics and Mechanics, Kazan Federal University,\\ Kremleovskaya str., 35, Kazan 420008, Russia}
\ead{ignatev\_yu@rambler.ru}
\begin{abstract}
In this article there are considered non-equilibrium cosmological scenarios with the assumption that scaling of particles interaction is restored in range of extra-high energies. On basis of energy-balance equation's exact solutions it is obtained the strong conclusion about fundamental unattainability of local thermodynamic equilibrium in the accelerated Universe. There are presented the results of numerical simulation of previously constructed strict mathematical model which describes thermodynamic equilibrium's establishment in the originally nonequilibrium cosmological ultrarelativistic plasma for the Universe with an arbitrary acceleration with the assumption that scaling of interactions of elementary particles is restored at energies above the unitary limit.  Limiting parametres of residual nonequilibrium distribution of extra-high energy relic particles are obtained.
The assumption about possibility of detection of "truly relic particles", which appeared at stage of early inflation, is put forward.
\end{abstract}



\section{Conditions for the local thermodynamic equilibrium of cosmological plasma}
One of the main statements of the standard cosmological scenario (SCS)\footnote{see, e.g.
\cite{S.Weinberg}} is an assumption of local thermodynamic equilibrium (LTE) of
cosmological plasma on the early stage of Universe expansion. As is known, to let
LTE establish in the statistical system, it is required that the effective time between
particles collisions, $\tau_{eff}$, is small as compared to the typical time scale of
system evolution. In the cosmological situation such scale is Universe age, and more
precisely, inverse value of the scale factor's $a(t)$ logarithmical derivative.
This brings us to the following well-known %
LTE condition in the expanding ultrarelativistic cosmological plasma\footnote{We choose  Planck system of units $G=\hbar=c=1$.}:
\begin{equation}\label{LTE1}
\frac{\dot{a}}{a}\tau_{eff}\ll 1, \Rightarrow \frac{\dot{a}}{a}\frac{1}{n(t)\sigma_{tot}}\ll 1,
\end{equation}
where $a(t)$ is a scale factor, $\dot{a}\equiv da/dt$, $n(t)$ is a particles number density, $\sigma_{tot}$ is a total cross-section of particles scattering in pair collisions.

\subsection{Kinematics of four-particles collisions}
Four-particles reactions of type
\begin{equation}\label{4particles}
a+b\rightarrow c+d.
\end{equation}
are fully described by two kinematic invariants, $s$ and $t$
(see, e.g. \cite{Pilk}):
\begin{equation}\label{s}
s=(p_a+p_b)^2\equiv (p_a+p_b,p_a+p_b),
\end{equation}
colliding particles' square energy in the center-of-mass system, and
\begin{equation}\label{Yu3}
t=(p_c-p_a)^2=(p_b-p_d)^2.
\end{equation}
the relativistic square of the transmitted momentum:\footnote{Author hopes that reader will not be confused by the coincidence of denotations: t  is a time in Friedmann metric, s is Friedmann metric interval, simultaneously t, s are
kinematic invariants. These denotations are standard and were not considered necessary to change.}. Next, $(p,q)\equiv g_{ik}p^iq^k$ is a scalar product of vectors $p,q$ relative to metric $g$, $a,b$ are particles indexes, $i,k=\overline{1,4}$; $\sqrt{s}$ is an energy of colliding particles in the center-of-mass system (CMS).

Invariant scattering amplitudes are determined as a result of
invariant scattering amplitude averaged by ``$c$'' and ``$d$'' particles' 
states at that are turned out to be dependent only on these two
invariants:
\begin{equation}\label{Yu4}
\overline{|M_{FJ}|^2}=|F(s,t)|^2,
\end{equation}
where $S_i$ are particles spins. Total cross-section of the reaction (\ref{4particles})  is determined by means of the invariant amplitude $F(s,t)$:
\begin{equation}\label{Yu5}
\sigma_{tot}= \frac{1}{16\pi\lambda(s,m^2_a,m^2_b)}\int\limits_{t_{min}}^{t_{max}} dt
|F(s,t)|^2,
\end{equation}
where $m_i$ are particles rest masses, $\lambda$ is a triangle function:
$$
\lambda^2(a,b,c)=a^2+b^2+c^2-2ab-2ac-2bc,$$
è
\begin{equation}
t^{\rm max}_{\rm min}=\frac{4}{s}
[ (m^2_c-m^2_a+m^2_b-m^2_d)^2 -(\sqrt{\lambda} \mp \sqrt{\lambda'})^2],\nonumber
\end{equation}
where the following denotations are introduced for the simplicity:
$$\lambda=\lambda(s,m^2_a,m^2_b);\; \lambda'=\lambda(s,m^2_c,m^2_d).$$
In the ultrarelativistic limit
\begin{equation}\label{Yu6}
\frac{p_i}{m_i} \to\infty
\end{equation}
we have:
\begin{eqnarray}\label{Yu6_1}
s\rightarrow 2(p_a,p_b);& t\rightarrow -2(p_a,p_b);& \lambda\to s^2;\nonumber
\\[12pt]
\label{t_ultra}
t_{min}=\rightarrow -s;& t_{max}\rightarrow 0;&
\frac{s}{m^2_i}\to\infty,\nonumber
\end{eqnarray}
and formula (\ref{Yu5}) is significantly simplified by the introduction of the dimensionless variable:
\begin{equation}\label{Yu7}
x=-\frac{t}{s}:
\end{equation}
\begin{equation}\label{Yu8}
\sigma_{tot}(s)=\frac{1}{16\pi s}\int\limits _{0}^1 dx|F(s,x)|^2.
\end{equation}
Thus, total cross-section of scattering depends only on kinematic invariant $s$ -
square energy of colliding particles in the center-of-mass system:
\begin{equation}\label{sigma(s)}
\sigma_{tot}=\sigma_{tot}(s).
\end{equation}
Exactly this dependency controls LTE establishment in the early Universe.

\subsection{Influence of $\sigma(s)$ on the process of LTE restoration}
In the space-flat Friedmann metric\footnote{and for any other Friedmann metric} which is considered in the article,
\begin{equation}\label{ds}
ds^2=dt^2-a^2(t)(dx^2+dy^2+dz^2)
\end{equation}
integral of motion is a {\it conformal momentum} (see e.g. \cite{Yu_1982})
\begin{equation}\label{ap}
\tilde{p}=a(t)p={\rm const},
\end{equation}
where $p=\sqrt{-g_{\alpha\beta}p^\alpha p^\beta}$ is a physically three-dimensional momentum %
($\alpha,\beta=\overline{1,3}$) $a$ is a scale factor. Thus, for ultrarelativistic particles
\begin{equation}\label{s(t)}
s\sim p^2\sim a^{-2}.
\end{equation}
Assuming power dependence of total particles cross-section  on $s$
\begin{equation}\label{sigma}
\sigma_{tot} \sim s^\nu
\end{equation}
and {\it barotropic} summary equation of matter state ${\rm p}=\varkappa\varepsilon$,
where ${\rm p}$ is a summary pressure \footnote{unlike momentum, $p$, pressure is highlighted by Roman type, ${\rm p}$.}, $\varepsilon$ is a summary energy density, $\varkappa$ is a {\it barotrope factor}, under the assumption of total particles number conservation $n(t)a^{3}(t)={\rm const}$ we arrive to the following conclusion \cite{Yu_1986}.

\begin{thm}\label{stat1}
At fulfillment of the condition
\begin{equation}\label{Yus_7}
4\nu+3(1-\varkappa)>0;\quad (\varkappa\not=-1),
\end{equation}
LTE is maintained on early stages of expansion and is broken on the late ones, i.e., at:
\begin{equation}\label{Yus_8}
\nu > -\frac{3}{4}(1-\varkappa)  \Rightarrow \; LTE\; \mbox{\rm at}\; t<t_0,
\end{equation}
and at fulfillment of the inverse to (\ref{Yus_8}) condition LTE is broken on early stages and is restored at late ones $t>t_0$.
\end{thm}

In particular,\\
$1^0$. in case of ultrarelativistic equation of state $\varkappa=1/3$ we obtain from
(\ref{Yus_8}) the condition of LTE existence on early stages of expansion
\cite{Yu_1986} ---
\begin{equation}\label{Yus_8a}
\nu > -\frac{1}{2} \Rightarrow \;LTE\; \mbox{\rm at}\; t<t_0,\quad p=\frac{1}{3}\varepsilon;
\end{equation}
$2^0$. in case of utmost rigid equation of state  $\varkappa=1$ condition
of LTE maintenance on early stages and breaking on late stages is equivalent to the condition
\begin{equation}\label{Yus_8b}
\nu > 0 \Rightarrow \;LTE\; \mbox{\rm at}\; t<t_0,\quad(p=\varepsilon)
\end{equation}
(hence it follows that at constant cross-section of scattering $\nu=0$ in case of utmost rigid equation
of state time is completely excluded from the condition of LTE \cite{Yu_1986}, -
at this stage of expansion in the Universe LTE either exists at all times or always is absent.);\\
$3^0$. in case of inflation  $\varkappa=-1$ at fulfillment of the condition
\begin{equation}\label{Yus_9}
\nu >-\frac{3}{2}
\end{equation}
LTE is maintained at early stages ($t<t_0$) and is broken on late ones ($t>t_0$).
The last is obviously true at conservation of the total particles number at the inflationary stage of expansion.

\section{Unified asymptotic cross-section of scattering}
\subsection{Unitarity and unitary limit}
For the investigation of the early Universe processes' kinetics it is required to know an asymptotic behavior of the invariant amplitudes$F(s,t)$ in the limit (\ref{Yu6}).  Modern experimental possibilities are limited by values $\sqrt{s}$  of the order of several Tev. It would be incautious to bear on one or another field interaction model for the forecasting of the asymptotic behavior of the scattering cross-section in range of extra-high energies of the order of
$10^{11}\div 10^{16}$ Tev. At modern conditions it looks more reasonable to bear on the conclusions of the asymptotic theory of  $S$-matrix, which are obtained on the basis of fundamental laws of unitarity, causality, scale invariance etc. The unitarity of the $S$-matrix leads to the well-known asymptotic relation (see e.g. \cite{Okun}):
\begin{equation}\label{As1}
\left.\frac{d\sigma}{dt}\right|_{s\to\infty}\sim \frac{1}{s^2}
\end{equation}
at value of $s$ above the unitary limit, i.e under the condition (\ref{Yu6}), if by $m_i$
we understand masses of all intermediate particles. However then from (\ref{Yu8}) it follows:
\begin{equation}\label{As2}
F(s,1)|_{s\to\infty}\sim \mbox{Const}. \end{equation}
Conception of the  {\it unitary limit} was first introduced by L.D. Landau in 1940 for vector mesons \cite{Land_Unit}. Following that article, by unitary limit  energy we  understand such a critical energy, above the value of which the growth of the effective cross-section of scattering is stopped and its behavior obeys the condition of unitarity.  For example, for the standard $\nu e$ - scattering the energy of the unitary limit $E_u=\sqrt{s_u}$ is (see e.g. \cite{Okun}) $\sqrt{\sqrt{2}\pi/G_\nu}$  $\approx600$ Gev, where $G_\nu$ is an electroweak interaction constant.

\subsection{Asymptotic behavior of particles scattering cross-sections in range of extra-high energies}
On the basis of the axiomatic theory of the  $S$ - matrix in 60's there were obtained strict limitations on the asymptotic behavior of the total cross-sections of scattering and its invariant amplitudes:
\begin{equation}\label{As3}
\frac{C_1}{s^2\ln s}< \sigma_{tot}(s)<C_2\ln^2 s,
\end{equation}
where $C_1,C_2$ are unknown constants. The upper limit  (\ref{As3})
has been determined in articles \cite{asw1}, \cite{asw2}, \cite{asw3}, the lower one - in articles
\cite{asw4}, \cite{asw5}, see also a review in the book \cite{asw6}.
Let us also point out limitations on the invariant amplitudes of scattering
 \cite{asw6}:
\begin{equation}\label{As4}
|F(s,t)|\leq |F(s,0)|;
\end{equation}
\begin{equation}\label{As5}
C'_1 < |F(s,0)|<C'_2 s\ln^2s.
\end{equation}

Thus invariant amplitudes of scattering in the limit (\ref{Yu6})
must be functions of only one variable $x=-t/s$, i.e.:
\begin{equation}\label{Yu15a}
|F(s,t)|=|F(x)|, \; (s \to\infty).
\end{equation}
However then as a result of (\ref{Yu8})
\begin{equation}\label{Yu15b}
\sigma_{tot}(s)=\frac{1}{16\pi s}\int\limits_0^1 dx
|F(x)|^2=\frac{\mbox{Const}}{s}
\end{equation}
--- the total cross-section's behavior is the same as for the electromagnetic interactions, i.e. at extra-high energies, scaling is  restored.

Scaling asymptotics of scattering (\ref{Yu15b}) lies strictly in the middle of possible extreme asymptotics of total scattering cross-section (\ref{As3}). Moreover, if  (\ref{Yu15b}) is fulfilled, relations (\ref{As1}) and (\ref{As2}) obtained on the basis of the axiomatic theory of the $S$ - matrix are also automatically fulfilled

For the lepton-hadron interaction a suggestion of scaling existence has been put forward in articles \cite{scal1}, \cite{scal2}, \cite{scal3}. In particular, for the total cross-section of the reaction "$e+e^+\rightarrow \mbox{hadrons}$" an expression has been obtained:
$$\sigma_{tot}=\frac{4\pi\alpha^2}{3s}\sum e^2_i,$$
where $\alpha$ is a fine structure constant, $e_i$ are charges of fundamental fermion fields. The data obtained on the Stanford collider, confirmed scaling existence for these interactions. For the gravitational interactions scaling, apparently, should be restored at extra-high energies in consequence of the scale invariance of the gravitational interactions in WKB-approach \cite{scal4}. There can be made a lot of similar examples which are reliable established facts.

\subsection{Universal asymptotic cross-section of \\ scattering}

Let us hereinafter suggest the existence of scaling at energies above the unitary limit $s\to\infty$. Then the question about the meaning of the constant in formula (\ref{Yu15b}) and also about the logarithmic refinement of this constant arises. This value can be estimated from the following simple considerations.  First, let $m$ be a rest mass of the colliding particles. Since $\sqrt{s}$ is an energy of the interacting particles in the center-of-mass system, the minimum value $\sqrt{s}$ for the four-particles reactions with the particles of mass
$m$ is:
\begin{equation}\label{s=4m^2}
\sqrt{s_{\rm min}}=2m \Rightarrow s_{\rm min}=4m^2.
\end{equation}
Next,  if the idea about unification of all interactions on Planck
energy scales $E_{pl}=m_{pl}=1$ is correct, then at $s\sim 1$ all
four-particle interactions must be described by the single
cross-section of scattering, that is formed of three fundamental
constants $G,\hbar,c$, i.e., in the chosen system of units it should
be:
\begin{equation}\label{15c}
\sigma|_{s\sim 1}=2\pi l^2_{pl}\Rightarrow
\sigma(4)=\frac{8\pi}{s_{pl}}\quad (=2\pi),
\end{equation}
where:
\begin{equation}\label{s_pl}
s_{pl}=4m^2_{pl}=4
\end{equation}
is a Planck value of the kinematic invariant $s$ corresponding to
two colliding plankeons with mass $m_{pl}$ and Compton scale
$l_{pl}$:
\begin{equation}\label{m_pl}
m_{pl}=\sqrt{\frac{\hbar c}{G}}\; (=1),\quad
l_{pl}=\sqrt{\frac{G\hbar}{c^3}}\; (=1).
\end{equation}

However, in order to decrease to such value on Planck  energy
scales, starting from values of order $\sigma_T=8\pi\alpha^2/3m^2_e$
($m_e$ - electron mass, $\sigma_T$ - Thompson cross-section of
scattering) for the electromagnetic interactions, i.e. at $s \sim
4m^2_e$, cross-section of scattering should decrease in inverse
proportion to $s$, i.e. again by scaling law. Let us note that {\bf this
fact is one more independent argument in favor of existence of
scaling in range of extra-high energies}. Refining this dependency
logarithmically we introduce the {\it universal asymptotic
cross-section of scattering} (ACS), which first was proposed in
articles \cite{UACS} (1984), \cite{Yu_1986}, (see also \cite{LTE}):
\begin{equation}\label{Yu15d}
\sigma_0(s)=\frac{8\pi\beta}{sL(s)},
\end{equation}
where $\beta\sim 1$,  $L(s)$ is a logarithmical factor:
\begin{equation}\label{yu15e}
L(s)=1+\ln^2\left(1+\frac{s_0}{s}\right)> 1,
\end{equation}
which is a monotone decreasing function of the kinematic invariant
$s$: $d L/ds<0,$ and $s_0=4$ is a squared total energy of two colliding Planck masses
so that on Planck energy scales:
\begin{equation}\label{Lambda(s0)}
L(s_0)\simeq1,
\end{equation}
at that on Compton scales of energy, i.e. at $s=m^2_e$:
\begin{equation}\label{alpha^2}
\frac{1}{\sqrt{L(m^2_e)}}\approx \frac{1}{102}\simeq
\alpha\approx\frac{1}{137},
\end{equation}
where  $\alpha=1/137$ is a fine structure constant.

Relation (\ref{alpha^2}) allows to consider value $1/\sqrt{L(s)}$ as
logarithmically changing effective constant of interaction, which,
in its turn, realizes the ideology of running interaction constants
of standard theories of $SU(5)$ - type fundamental interactions.

Introduced by formual (\ref{Yu15d}), cross-section of scattering
$\sigma_0$, ACS, possesses a number of remarkable features (see also Figure \ref{uacs_fig}):\\[10pt]
$1^o$. ACS is formed with a use of only fundamental constants $G, \hbar, c$;\\
$2^o$. ACS behaves itself so that its values lie strictly in the
middle of possible extreme limits of the asymptotic behavior of
cross-section (\ref{As3}), which were established by means of the
asymptotic theory of the
$S$-matrix;\\
$3^o$. With logarithmic
accuracy ACS is a scaling cross-section of scattering ;\\
$4^o$. ACS with a remarkable accuracy coincides with cross-sections
of all known fundamental processes on the corresponding energy
scales, starting from the electromagnetic and finishing with
gravitational ones on the huge range of energy values (from $m_e$ to
$10^{22}m_e$) --- the values of the first kinematic invariant at
that changes by an order of of 44!! (see Figure \ref{uacs_fig}).
\\[10pt]
In the previous works there were given the arguments in favor of scaling behavior of the
scattering cross-section in range {\it of energy above the unitary limit}. Among them there was provided the comparison of ACS with the specific cross-sections of scattering for certain specific four-particle reactions at the corresponding energy ranges. However,
during the discussions with experts in the quantum field theory the unacceptance of
the statement about scaling behavior of the scattering cross-section is often found out. The Author suggests that this unacceptance is caused,
first of all, by quite limited range of energies for which the calculations of
concrete scattering cross-sections are carried out, and second, by the ``look from below'' on quantum procedures of calculation of scattering
cross-sections in sense
that energy is considered below the unitary limit. For elimination of this misunder\-stan\-ding" in this article Author represents the comparison of the asymptotic scattering cross-section's values
(\ref{Yu15d}) with the known quantum four-partial processes in the gra\-phic format (Figure  \ref{uacs_fig}).
\centerline{\includegraphics[width=12cm]{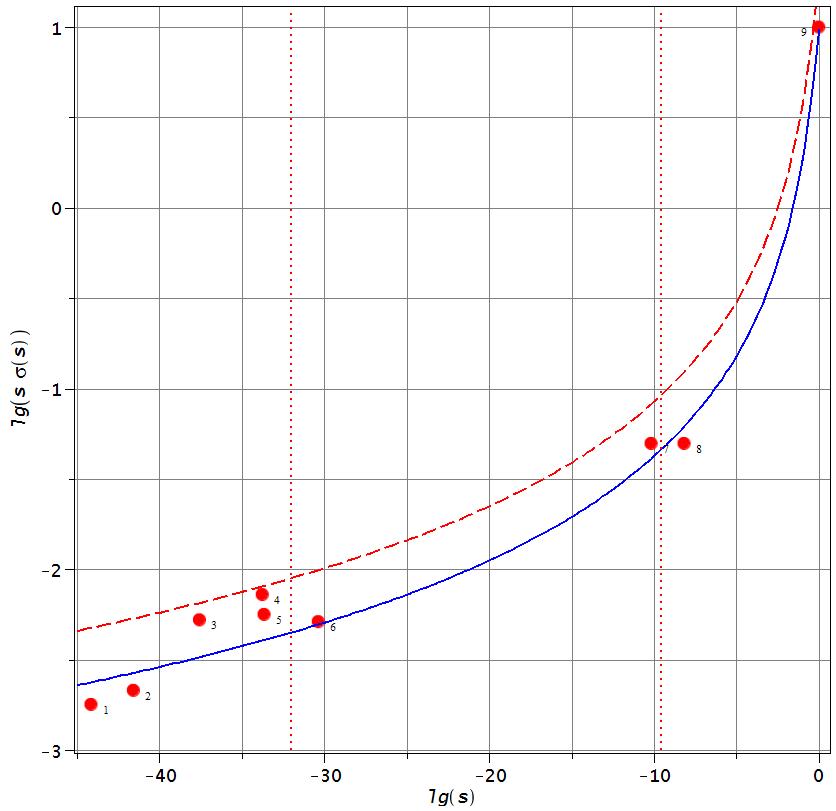}}
\refstepcounter{figure}{Figure \thefigure.}\label{uacs_fig}
\hskip 12pt {\small Comparison of the universal cross-section of
scattering (\ref{Yu15d}) at factor $\beta=1$ with the well-known
cross-sections of fundamental processes is noted by bold line. Dotted line
corresponds to the graph of universal cross-section of scattering at
factor  $\beta=2$. On the abscissa axis are laid values of the
common logarithm of the first kinematic invariant $\lg s$ in Planck
units; along the ordinate axis are laid values of the common
logarithm of the dimensionless invariant, $\lg s\sigma(s)$.  1 --
Thompson scattering, 2 -- Compton scattering on electrons at
$E_{SCM}=10$ Mev, 3 -- Compton scattering of electrons at
$E_{SCM}=1$ Gev, 4 -- electroweak interaction with participation of:
$W$ - bosons, 5 -- with participation of $Z$ - bosons, 6 -- with
participation of H-bosons at energy of the order of  7 Tev
($\sigma\sim 10$fb); 7 -- $SU(5)$ - interaction at mass of the
superheavy X-bosons $10^{15}$ Gev, 8 -- $10^{14}$ Gev; 9 --
gravitational interaction on Planck scales. Vertical dotted lines
correspond to energy values of the unitary limit for $SU(2)\times
SU(1)$ - interac\-tions, $E_u=617$ Gev, and SU(5) - interac\-tions,
$E_u\sim10^{14}$ Gev.} \vskip 10pt

These remarkable properties of ACS hardly can be accidental and
allow us further use ACS as a reliable formula for the asymptotic
value of scattering cross-sections for all interactions.

\section{Kinetic equations for superthermal particles}
\subsection{Simplification of relativistic collision integral}
Process of thermodynamic equilibrium establishment is described by
relativistic kinetic equations. In paper \cite{Yu_1982} it
is shown that relativistic kinetic equations are conformally
invariant in ultrarelativistic limit at existence of scaling of
interactions. This fact is the basis for the assertion that at least
in the ultrarelativistic Universe LTE could be broken. So, we
consider homogenous isotropic distributions of particles in
Friedmann metric (\ref{ds}). Such distributions are described by
functions:
\begin{equation}\label{f(t,p)}
f_a(x^i,p^k)=f_a(t,p). \end{equation}
Relativistic kinetic equations relatively to homogenous isotropic
distributions  (\ref{f(t,p)}) take form (details see in
\cite{Yukin3}, \cite{Yukin3_1}, \cite{Yukin2}, \cite{Yukin4},
\cite{Yukin1}):
\begin{equation}\label{II.4}
\frac{\partial f_a}{\partial t}-\frac{\dot{a}}{a} p\frac{\partial f_a}{\partial
p}=\frac{1}{\sqrt{m_a^2+p^2}}\sum\limits_{b,c,d}^{} J_{ab}(t,p),\end{equation}
where  $J_{ab}(t,p)$ -is an integral of 4-particle reactions
\cite{Yukin1}, \cite{Yukin2}:
\begin{eqnarray}
J_{ab}(t,p) =(2\pi)^4 \int d\pi_b d\pi_c d\pi_d \delta^{(4)} (p_a+p_b-p_c-p_d)\\
\times  [(1\pm f_a)(1\pm f_b)|f_cf_d \overline{|M_{cd\to ab}|^2}- (1\pm f_c)(1\pm
f_d)f_af_b|\overline{M_{ab\to cd}|^2}\,],\label{II.7} \nonumber
\end{eqnarray}
signs $\pm$ correspond to bosons ($+$) and fermions ($-$), $M_{i\to
f}$ are invariant scattering amplitudes (underscore means averaging
by particles' polarization states), $d\pi_a$ is a normalized volume
element of momentum space of  $a$-particle:
\begin{equation}\label{II.8}
d\pi_a=\sqrt{-g}\frac{\rho_a dp^1dp^2dp^3}{(2\pi)^3p_4},
\end{equation}
$\rho_a$ - degeneration factor.

Let us simplify the integral of 4-particle interactions
(\ref{II.7}), using properties of distributions  $f_a(t,p)$
isotropy. For the fulfillment of two inner integrations by momentum
variables we pass to the local center-of-mass system, where
integration is carried out very simply. After the inverse Lorentz
transform and proceed to the spherical coordinate system in the
momentum space in the ultrarelativistic limit (\ref{Yu6}) we find
(\cite{Yu_1986}, \cite{LTE2}):
\begin{eqnarray}\label{II.12}
J_{ab}(p)\!\!=\!\!-{\displaystyle\frac{2S_b+1}{(2\pi)^3p}}{\displaystyle\int\limits_0^\infty}\! dq {\displaystyle\int\limits_{0}^{4pq}\!\!\frac{ds}{16\pi}}{\displaystyle\int\limits_{0}^1} dx
|F(x,s)|^2 {\displaystyle\int\limits_0^{2\pi}}\!\!d\varphi\times\nonumber\\
\{f_a(p)f_b(q)[1\pm f_c(p-\Delta)][1\pm f_d(q+\Delta)]\nonumber\\
-f_c(p-\!\!\Delta)f_d(q+\!\!\Delta)[1\pm\!\! f_a(p)][1\pm\!\! f_b(q)]\},
\end{eqnarray}
where $x=-t/s$ dimensionless variable (\ref{Yu7}),
and
\begin{equation}\label{II.14}
\Delta=x(p-q)-\cos\varphi \sqrt{x(1-x)(4pq-s)}.
\end{equation}
\subsection{Relativistic kinetic equations in terms of conformally corresponding space \label{KinEq_Conf}}
Taking into account the fact that variable $\tilde{p}$ (\ref{ap}) is
an integral of motion in Friedmann metric and at that for any
function $\Psi(t,p)$ holds the relation \cite{Yukin3}:
\begin{equation}\label{II.30}
\frac{\partial \Psi(t,p)}{\partial t}-\frac{\dot{a}}{a}p\frac{\partial
\Psi(t,p)}{\partial p}=\frac{\partial \Psi(t,\tilde{p})}{\partial t},
\end{equation}
we transform the kinetic equations for the homogenous isotropic
distributions to the form:
\begin{equation}\label{Kin_Frid}
\frac{\partial f_a}{\partial \eta}=\frac{a}{\sqrt{m_a^2+p^2}}\sum\limits_{b,c,d}^{}
J_{ab}(\eta,p) ,
\end{equation}
where it is necessary to substitute $p=\tilde{p}/a$.

Let us note that from the other hand, transformation to varaible
(\ref{ap}), $\tilde{p}$, is practically a conformal transformation
to the homogenous static space
$$ds^2=a^2 ds_0^2=a^2(d\eta^2-dl^2),$$
where physical component of the momentum, $p$, is transformed by the
law:
\begin{equation}\label{phys_p_conf} p=\frac{\tilde{p}}{a}.
\end{equation}
Thus, the momentum variable (\ref{ap}), $\tilde{p}$, is an absolute
magnitude of the physical momentum in the conformally corresponding
static space of constant curvature\footnote{in the considered case -
in the Minkovsky space}, and $\eta$ is a time variable in this
space.

Particles number densities, $n(\eta)$, and their energy densities,
$\varepsilon(\eta)$,  relatively to the isotropic particles
distribution, $f(\eta,p)$ are determined by formulas):
\begin{equation}
\label{n_no} n(\eta)={\displaystyle \frac{\rho}{2\pi^2}\int\limits_0^\infty} p^2 f(\eta,p)dp =
{\displaystyle\frac{\rho}{2\pi^2a^3}\int\limits_0^\infty} \tilde{p}^2
 f(\eta,\tilde{p})d\tilde{p};
  \end{equation}
 \begin{equation}\begin{array}{l}
\label{E_no} \varepsilon(\eta)={\displaystyle\frac{\rho}{2\pi^2}\int\limits_0^\infty}
\sqrt{m^2+p^2}p^2 f(\eta,p)dp =\\ = {\displaystyle\frac{\rho}{2\pi^2a^3}\int\limits_0^\infty}\tilde{p}^2
\sqrt{m^2+\tilde{p}^2/a^2} f(\eta,\tilde{p})d\tilde{p}.
\end{array}
\end{equation}
In connection with this it is convenient to introduce conformal
particles number densities, $\tilde{n}(\eta)$, and for the
ultrarelativistic particles also their energy densities,
$\tilde{\varepsilon}(\eta)$:
\begin{eqnarray}
\label{n_conf} \tilde{n}(\eta)= {\displaystyle\frac{\rho}{2\pi^2}\int\limits_0^\infty} \tilde{p}^2
f(\eta,\tilde{p})d\tilde{p};\\
\label{E_conf} \tilde{\varepsilon}(\eta)={\displaystyle\frac{\rho}{2\pi^2}\int\limits_0^\infty}
\sqrt{m^2+p^2}p^2 f(\eta,p)dp
 \approx  {\displaystyle\frac{\rho}{2\pi^2}\int\limits_0^\infty}\tilde{p}^3
f(\eta,\tilde{p})d\tilde{p}.
\end{eqnarray}
Then two relations are held:
\begin{eqnarray}
\label{nn_conf} \tilde{n}(\eta)&=& n(\eta)a^3(\eta);\\
\label{ee_conf} \tilde{\varepsilon}(\eta) & \approx & \varepsilon(\eta)a^4(\eta); \qquad
(p/m\to\infty),
\end{eqnarray}
from which the first one is strictly fulfilled and the second is
fulfilled asymptotically in the ultrarelativistic limit.

\subsection{Collision integral for weak deviation of distributions from the equilibrium}%
Let us first investigate weak breakdown of the thermodynamic
equilibrium in the hot model, when the main part of particles
 $n_{e}(t)$, lays in the state of thermal equilibrium
and only for the small part, $n_{ne}(t)$, --
\begin{equation}\label{II.15}
n_{ne}(t)\ll n_e(t)
\end{equation}
thermal equilibrium is broken (see Figure \ref{Fig_df}).
Hence\-forth in this article we suppose that distribution functions
differ slightly from the equilibrium ones in range of energy small
values, under the certain unitary limit, $p=p_0$ (or $T=T_0$), below
which scaling is absent can freatly differ at energies above the
unitary limit:
\begin{equation}\label{II.16}
\!\!\!f_a(p)\approx\left\{\begin{array}{ll}%
\!\!\!f^0_a=\left[\exp(\frac{-\mu_a +E_a(p)}{T})
\pm 1\right]^{-1}, & p< p_0;\\
 & \\
\!\!\!\Delta f_a(p); \;f_a^0(p)\ll \Delta f_a(p)\ll 1, & p>p_0,\\
& \\
\end{array}\right.
\end{equation}
where $\mu_a(t)$ are chemical potentials, $T(t)$ is a temperature of
the equilibrium component of plasma. Thus, in range $p>p_0$ it can
be observed anomalously great number of particles as compared to the
equilibrium one, simultaneously small (see (\ref{II.15})) as
compared to the total number of equilibrium particles.

Let us investigate the process of distribution $f_a(p)$ relaxation
to the equilibrium $f^0_a(p)$. Problem in this statement for the
particular case of initial distribution $f(t=0,p)$ has been solved
earlier in articles \cite{UACS}, \cite{Yu_1986}, \cite{LTE2}. Here
we give the general solution of this problem. As we will see further
the cosmological plasma at that can formally be considered as a
two-component system of equilibrium part with distribution
$f_a^0(t,p)$, and non-equilibrium, {\it superthermal} part with
distribution $\delta f_a(t,p)=\Psi(t,p)$. Particles number in
non-equilibrium component at that is small, but its energy density,
generally speaking, is random. Let us investigate the collision
integral (\ref{II.12}) in range
\begin{equation}\label{II.17}p\geq p_0\gg T.
\end{equation}
\centerline{\includegraphics[width=10cm]{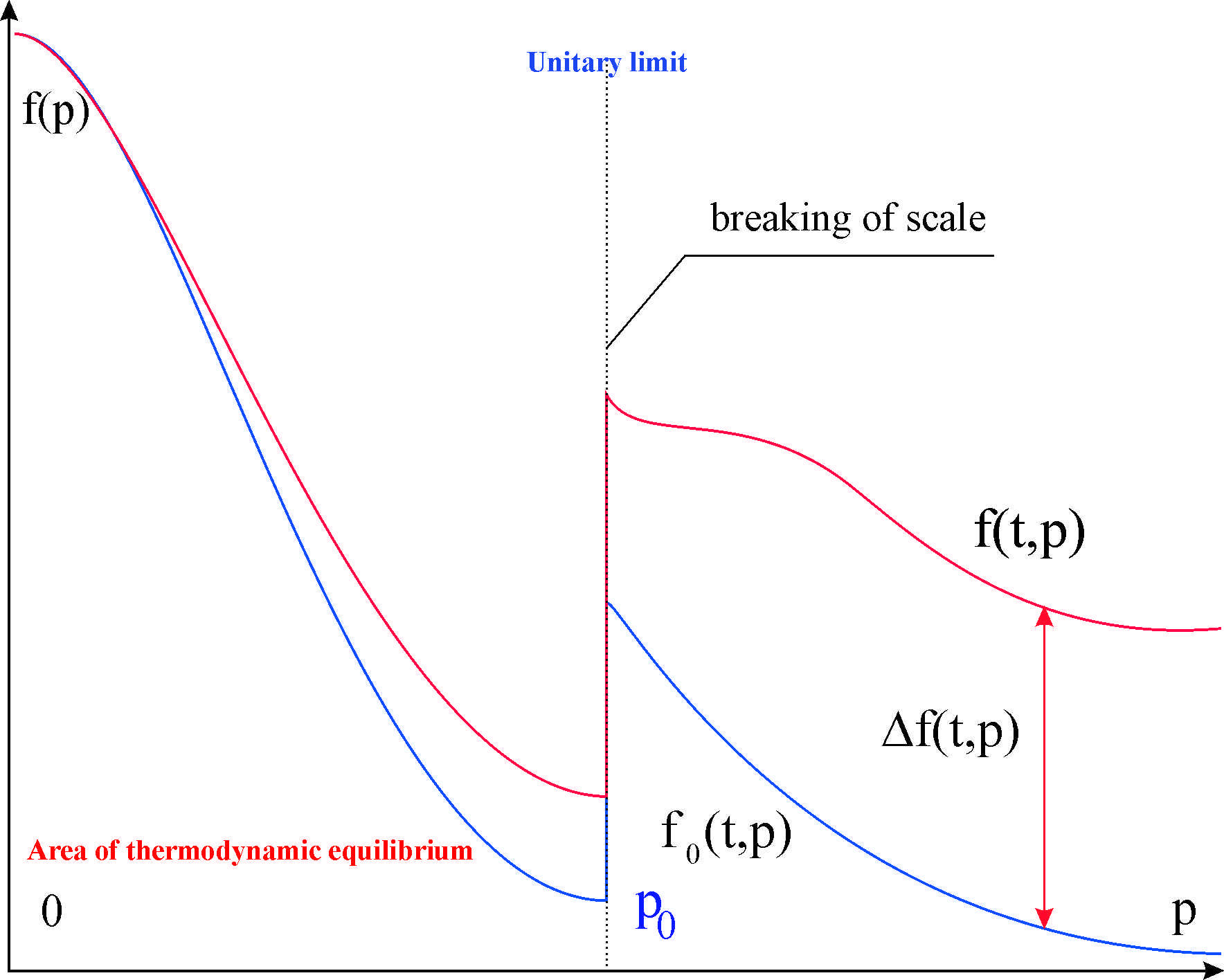}}
\refstepcounter{figure}{\bf Figure \thefigure.}\label{Fig_df} \hskip
12pt {\small Schematic representation of distribution function's
deviation from the equilibrium.} \vskip 12pt

In consequence of inequality (\ref{II.16}) we can neglect
superthermal particles' collisions between themselves in this range,
confining ourselves to account of superthermal particles scattering
on equilibrium particles. Therefore in collision integral the value
of one of the momentums, either $p'=p-\Delta$ or $q'=q+\Delta$
should be found in thermal range, the second one's - in superthermal
range above the unitary limit. Outside of this range the integrand
value of the collision integral is extremely small. As a result of
this fact we can neglect the second member  in curly brackets
(\ref{II.12}) since it can compete with the first one only in
asymptotically small variation ranges of variables $x$ and
$\varphi$: $x(1-x)\lesssim T/p\to 0$. Statistical factors of form
$[1\pm f_a(p')]$ in the first member of integral (\ref{II.12}) can
appreciably differ from one only in range of thermal momentum
values. As a result in the examined range of momentum values the
collision integral (\ref{II.12}) can be written in form \cite{LTE2}:
\begin{equation} \label{II.18}
\hspace{-40pt}\left.J_{ab\leftrightarrow cd}(p)\right|_{p\geq p_0}=\frac{(2S_b+1)\Delta
f_a(p)}{(2\pi)^3p}
\int\limits_0^\infty \frac{qf_b^0(q)dq}{\sqrt{m^2_b+q^2}}
\int\limits_{2p(q^4-q)}^{2p(q^4+q)}\frac{ds}{16\pi}\int\limits_0^1 dx|F(x,s)|^2.
\end{equation}
Using here the definition of the total cross-section of scattering
(\ref{II.12}), we obtain from (\ref{II.18}):
\begin{equation}\label{II.19}
\left.J_{ab\leftrightarrow cd}(p)\right|_{p\geq
p_0}=\frac{(2S_b+1)\Delta f_a(p)}{(2\pi)^3p}
\int\limits_0^\infty
\frac{qf_b^0(q)dq}{\sqrt{m^2_b+q^2}}
\int\limits_{2p(q^4-q)}^{2p(q^4+q)}\!\!\!\!\!\sigma_{tot}s(s)ds.
\end{equation}
Finally, substituting into the inner integral the expression for
$\sigma_{tot}$ in form of ACS, (\ref{Yu15d}), carrying out
integration with logarithmical accuracy and summing up the obtained
expression by all reactions channels we find:
\begin{equation}\label{II.20}
\left.J_a(p)\right|_{p\geq p_0}=-\Delta
f_a(p)\sum\limits_b\frac{4(2S_b+1)\nu_{ab}}{\pi} \int\limits_0^\infty \frac{q^2f^0_b(q)}{\sqrt{m^2_b+q^2}}\frac{dq}{L(\bar{s})},
\end{equation}
where $$\bar{s}=\frac{1}{2}pq^4,$$ $\nu_{ab}$ is a number of
channels of the reactions, in which can participate the $a$-sort
particle.

Let us calculate the values of integral (\ref{II.20}) in the extreme
cases.
\subsection{Expressions for equilibrium densities}
Let us write out expressions for macroscopic densities relative to
equilibrium distributions $f^0_a(p)$ (\ref{n_no}), (\ref{E_no}),
$\stackrel{0}{n}$,  and energy $\stackrel{0}{\varepsilon}$. In case
of massless particles gas ($\mu=0$) we obtain (see e.g.
\cite{Land}):
\begin{equation}\label{n_0}
\stackrel{0}{n}=\frac{\rho}{2\pi^2}\int\limits_0^\infty
\frac{p^2dp}{e^{p/T}\pm 1}=\frac{\rho T^3}{\pi^2}g_n\zeta(3);
\end{equation}
\begin{equation}\label{E_0}
\stackrel{0}{\varepsilon}=\frac{\rho}{2\pi^2}\int\limits_0^\infty
\frac{p^3dp}{e^{p/T}\pm 1}=\frac{\rho\pi^2 T^4}{30}g_e,
\end{equation}
where $\rho$ is a number of independent spin polarizations of
particle ($\rho=2$ for photons and massless neutrinos), $g_a$ is a
statistical factor:
\begin{eqnarray}\label{g_a}
g_n=g_e=1\quad \mbox{for Bose particles};\nonumber \\ g_n=3/4,\; g_e=7/8
\quad \mbox{for Fermi particles}. \end{eqnarray}
sign ``+'' corresponds to fermions, ``-'' - to bosons, $\zeta(x)$ -
$\zeta$ is Riemann function.

The total energy density of massless particles is equal to:
\begin{equation}\label{E}
\varepsilon=\sum\limits_a \stackrel{0}{\varepsilon}_a=N \frac{\pi^2 T^4}{15},
\end{equation}
where
\begin{equation}\label{g_E}
N=\frac{1}{2}\left[\sum\limits_B (2S+1) + \frac{7}{8}\sum\limits_F (2S+1)\right]
\end{equation}
is an effective number of particle types ($S$ is a spin of
particle)\footnote{In field interaction models of type  SU(5) ${\cal
N}\sim 100\div 200$.}; summation is carried out by bosons (B) and
fermions (F) correspondingly. Let us introduce numbers of bosons'
and fermions', $N_B$ and $N_F$:
\begin{equation}\label{N_BF}
N_B=\frac{1}{2}\sum\limits_B (2S+1);\quad
N_F=\frac{1}{2}\sum\limits_F (2S+1) .
\end{equation}
Then:
\begin{equation}\label{N}
N=N_B+\frac{7}{8}N_F.
\end{equation}

For the non-relativistic particles gas:
\begin{equation}\label{n_n}
\stackrel{0}{n}_a\approx  \delta \stackrel{0}{n}_\gamma =\delta\frac{2T^3}{\pi^2}\zeta(3),\; \varepsilon_a\approx m_a n_a,
\end{equation}
where $\stackrel{0}{n}_\gamma$ is a number density of relict
photons,
\begin{equation}\label{delta}
\delta\sim 10^{-10}.
\end{equation}
It should be noted that possible average concentration of particles
of non-baryon kind of dark matter $\delta_{nb}$, at its density of
order of 25\% from the critical one, $\rho_c\approx
0.9\cdot10^{-29}$ g/cm$^3$, and estimated minimal mass of particle
of order of  50 Gev is even less than $1/\delta$ and is of order of
$0.5\cdot10^{-11}$.
\subsection{Scattering on non-relativistic particles}
If equilibrium particles of $b$-sort are non-relativistic, i.e.
$q\ll m_b$ then integral(\ref{II.20}) is reduced to the expression:
\begin{equation}\label{II.21}
\left.J_a(p)\right|_{p\geq p_0}=-32\pi^2\Delta
f_a(p)
\sum\limits_b\frac{n^0_b(t)}{m_b}{\displaystyle
\frac{\nu_{ab}}{1+\ln^2pm_b/2}}, \quad (m_b>T).
\end{equation}
\subsection{Scattering on ultrarelativistic particles}
If equilibrium particles of $b$-sort are ultrarelativistic, i.e.
$m_b\ll T$ and their chemical potential is small - $\mu_b\ll T$ then
calculating integral (\ref{II.20}) with respect to equilibrium
distribution (\ref{II.16}) we find:
\begin{equation}\label{II.22}
\left.J_a(p)\right|_{p\geq p_0}=-\frac{4\pi}{3}\frac{\tilde{
N}T^2(t)}{1+\ln^2Tp/2}\Delta f_a(p),\quad
(m_b\ll T, \; \mu_b\ll T),
\end{equation}
where
$$\tilde{N}\!\!=\!\!\frac{1}{2}\left[\sum\limits_B (2S+1)\!\!+\!\!
\frac{1}{2}\sum\limits_F (2S+1)\right]=N_B+\frac{1}{2}N_F.$$ 
Calculating the relation of investments of non-relativistic and
ultrarelativistic equilibrium particles to the collision integral we
obtain:
\begin{equation}\label{II.27}
\frac{J_{non}}{J_{ultra}} \sim \frac{24\pi n^0_b}{m_b T^2}= \zeta(3)\delta \frac{64 T(t)}{\pi
m_b}\sim 10^{-9} \frac{T}{m_b},
\end{equation}
- relation of investments is small at $T\ll 10^9m_b$ and decreases
with time. Therefore hence we neglect the investment of
non-relativistic particles into the collision integral.

\section{Construction and solution of the energy-balance equation}
Let us build the strict self-consistent mathematical model of thermal equilibrium establishment in the expanding Universe  at conditions of equilibrium weak violation in terms of smallness of
the number of %
non-equilibrium particles as compared to the number of equilibrium
ones (\ref{II.15}). Let us note that energy contained in the
non-equilibrium high-energy ``tail'' of the distribution 
$\Delta f_a(t,p)$ at that can be very high and even sufficiently
exceed the energy of the cosmological plasma's  equilibrium
component. Let us write out first main relations determining
cosmological plasma dynamics.
\subsection{Model of matter}%
As is known (see e.g. \cite{Land_Field}), Einstein equations in case
of isotropic homogenous cosmological model with zero
three-dimensional curvature are reduced to the system of two
ordinary differential equations of the first order:
\begin{equation}\label{IV.1}
\frac{\dot{a}^2}{a^2}=\frac{8\pi}{3}\varepsilon;
\end{equation}
\begin{equation}\label{IV.2}
\dot{\varepsilon}+3\frac{\dot{a}}{a}(\varepsilon+{\rm p}(\varepsilon))=0.
\end{equation}
Next:
\begin{equation}\label{IV.3}
\varepsilon=\varepsilon_p+\varepsilon_s;\quad {\rm p}={\rm p}_p+{\rm p}_s,
\end{equation}
where $\varepsilon_p,{\rm p}_p$ are the energy density and pressure
of the cosmological plasma, $\varepsilon_s,p_s$ are the energy
density and pressure of all possible fundamental fields, probably
scalar, which lead to the Universe acceleration.

An {\it invariant acceleration of the Universe}
\begin{equation}\label{IV.4}
\Omega=\frac{a\ddot{a}}{\dot{a}^2}
\end{equation}
is connected with the {\it effective barotrope coefficient} of
matter, $\varkappa\equiv {\rm p}/\varepsilon$, by the relation:
\begin{equation}\label{IV.5}
\Omega=-\frac{1}{2}(1+3\varkappa).
\end{equation}
Thus, on the acceleration stage  ($t>t_1$) it is:
\begin{equation}\label{IV.6}
0<\Omega<1, \Rightarrow -1 <\varkappa<-\frac{1}{3}.
\end{equation}
Before that moment it was  $\varkappa\in [1/3,-1/3]$. According to
(\ref{IV.1}) -- (\ref{IV.2}) the scale factor and the summary energy
density {\it at given constant barotrope coefficient} are changing
by law:
\begin{equation}\label{IV.7}
a=a_1 t^{2/3(\varkappa+1)};\; \varepsilon=\frac{1}{6\pi
(\varkappa+1)^2t^2}, \quad \varkappa+1\not=0
\end{equation}
Let us rewrite the relations (\ref{IV.7}) in more convenient form
using the relation (\ref{IV.5}):
\begin{equation}\label{IV.8}
a=a_1 t^{1/(1-\Omega)};\; \varepsilon=\frac{3}{8\pi
(1-\Omega)^2t^2}, \quad \Omega<1.
\end{equation}

Let us note that at any values of the invariant acceleration $\Omega
={\rm Const}\in [-1,1)$ the energy density is proportional to
$t^{-2}$.
\subsection{Basic model assumptions}

\begin{thm}\label{stat2}Let us accept the following assumptions within the model:\\
\noindent\underline{$1^o$}. \hskip 12pt Scaling character of
elementary particles' interactions in range of extra-high energies
and unification of all interparticle interactions on the basis of
fundamental constants $G,\hbar,c$ at energies above the unitary
limit, which leads us to the formula of the asymptotic cross-section
of scattering (\ref{Yu15d});\\
\noindent\underline{$2^o$}. \hskip 12pt {\it The minimality of
connectivity between the fundamental  macroscopic fields and
cosmological plasma}. This automatically means that the energy
conservation law (\ref{IV.8}) is held separately for the fundamental
fields and plasma:
\begin{eqnarray}\label{IV.10}
\dot{\varepsilon_s}+3\frac{\dot{a}}{a}(\varepsilon_s+{\rm p}_s(\varepsilon_s))=0,\\%
\label{IV.11}
\dot{\varepsilon_p}+3\frac{\dot{a}}{a}(\varepsilon_p+{\rm p}_p(\varepsilon_p))=0.
\end{eqnarray}
\noindent\underline{$3^o$}. \hskip 12pt {\it Ultrarelativistic
equation of cosmological plasma state} on the considered stage of
expansion:
\begin{equation}\label{IV.12}
p_p=\frac{1}{3}\varepsilon_p,\quad (\ref{IV.11}) \Rightarrow
\varepsilon_pa^4={\rm Const}.
\end{equation}
\noindent\underline{$4^o$}. \hskip 12pt  {\it Ultrarelativistic
start of the Universe}:
\begin{equation}\label{IV.13}
\lim_{t\to 0}\varkappa=\frac{1}{3}.
\end{equation}
\end{thm}

\subsection{Energy balance of the cosmological plasma}
The basis of the developed theory here is the  {\it energy-balance
equation} of the cosmological plasma, which is in fact the
conservation law of its energy. For the case of summary
ultrarelativistic state of the matter this theory has been developed
by author in articles \cite{Yu_1986}, \cite{LTE2}. Here we
generalize and specify the results of this theory for the case of
the arbitrary summary equation of matter state. From (\ref{IV.11})
with the account of (\ref{IV.12}) it right away follows:
\begin{equation}\label{IV.17}
\varepsilon_p a^4\equiv \tilde{\varepsilon}_p = {\rm Const},
\end{equation}
where $\tilde{\varepsilon}_p$ is a conformal energy density of the
cosmological plasma. Let us define this constant assuming according
to (\ref{IV.8}) on the initial ultrarelativistic expansion stage it
is:
\begin{equation}\label{IV.18}
\left. a(t)\right|_{t\to 0}=\sqrt{t}.
\end{equation}
Then for the conformal energy density of plasma assuming that
cosmological plasma is the single ultrarelativistic cimponent of
matter, we obtain
\begin{equation}\label{IV.19}
\tilde{\varepsilon}_p=\frac{3}{32\pi}.
\end{equation}
Let us further introduce the temperature $T_0(t)$ of the
cosmological plasma in the ideal Universe where on the given moment
of cosmological time $t$ all plasma is locally equilibrium. Thus,
energy density of this plasma is described by formula (\ref{E}) with
$N=N_0$ being an effective number of equilibrium particles types in
plasma with temperature $T_0$. Hence subject to (\ref{IV.19}) we
obtain the evolution law of plasma temperature in the expanding
Universe:
\begin{equation}\label{IV.21}
T_0(t)=\frac{1}{a(t)}\left(\frac{45}{32\pi^3 N_0}
\right)^{\frac{1}{4}}.
\end{equation}
Relative to the value of $N_0$ {\it the effective number of types of
particles}, lying in thermodynamic equilibrium, we will assume that
$N_0(t)$ is a slowly changing function of the cosmological time:
\begin{equation}\label{IV.22}
\dot{N}_0 t\ll 1.
\end{equation}

Let now $T(t)$ be the genuine temperature of the cosmological
plasma's equilibrium component, and $\Delta f_a(p,t)$ be the
distribution function of <<$a$>> -sort non-equilibrium particles of
plasma. Let us find energies of the equilibrium, $\varepsilon_e $,
and non-equilibrium, $\varepsilon_{ne}$, components:
\begin{eqnarray}\label{IV.23}%
\varepsilon_e = & {\displaystyle\frac{N\pi^2}{15}}T^4(t);\\
\label{IV.24}%
\varepsilon_{ne} = & {\displaystyle\frac{1}{2\pi^2}}\sum\limits_a (2S+1)
\int\limits_0^\infty p^3\Delta f_a(p,t)dp,
\end{eqnarray}
where $S$ is a spin of particles; $N(t)$ is an effective number of
equilibrium particle types in plasma with temperature $T(t)$.
Expressing further the scale factor via temperature $T_0(t)$ by
means of (\ref{IV.21}) and introducing the new {\it dimensionless}
conformal momentum variable $\tilde{p}$:\footnote{As opposed to the
momentum variable $p$, pressure is denoted by Roman type, -- ${\rm
p}$. }
\begin{equation}\label{IV.25}
p=\left(\frac{45}{32\pi^3}\right)^{\frac{1}{4}}\cdot\frac{\tilde{p}}{a(t)}=T_0(t)N_0^\frac{1}{4}\tilde{p},
\end{equation}
for the (\ref{IV.24}) we obtain:
\begin{equation}\label{IV.26_0}
\tilde{\varepsilon}_{ne}=\frac{45}{64\pi^5}\sum\limits_a (2S+1)
\int\limits_0^\infty \tilde{p}^3\Delta f_a(\tilde{p},t)d\tilde{p}.
\end{equation}
Next, from (\ref{IV.21}) and (\ref{IV.23}) for the conformal energy
density of plasma equilibrium component we obtain:
\begin{equation}\label{IV.26}
\tilde{\varepsilon}_e=\frac{3}{32\pi}y^4,
\end{equation}
where it is introduced the dimensionless function, $y(t)$ -- {\it
relative temperature} \cite{LTE2}:
\begin{equation}\label{IV.27}
y(t)=\frac{T(t)}{T_0(t)}\leqslant 1.
\end{equation}
From (\ref{IV.26}) we can obtain a ratio:
\begin{equation}\label{IV.28}
\sigma(t)\equiv y^4(t)\equiv
\frac{\varepsilon_e}{\varepsilon_p}\equiv
\frac{\tilde{\varepsilon}_e}{\tilde{\varepsilon}_e+\tilde{\varepsilon}_{ne}}.
\end{equation}

Thus, cosmological plasma's energy conservation law (\ref{IV.19})
with a use of relations (\ref{IV.24}) and (\ref{IV.26}) can be
rewritten in from
\begin{equation}\label{IV.29}
y^4+\frac{15}{2\pi^4}\sum\limits_a (2S+1) \int\limits_0^\infty
\tilde{p}^3\Delta f_a(\tilde{p},t)d\tilde{p}=1.
\end{equation}
The relation (\ref{IV.29}) is called {\it plasma energy-balance
equation}. It is obtained with a use of three model suggestions
--- $2^o$, $3^o$, $4^o$. Let us note that in author's previous articles this
basic relation of the mathematical model of
ther\-mo\-dy\-na\-mi\-cal equilibrium's restoration has been
obtained under more special suggestions. At given dependency of
nonequilibrium particles' distribution function on the temperature
of plas\-ma's equilibrium component and cosmological time the
energy-balance equation becomes a nonlinear integ\-ral equa\-tion
relative to equilibrium compo\-nent's tempe\-ra\-tu\-re. Therefore,
to obtain this equa\-tion in the explicit form it is necessary to
solve the kinetic equation for the nonequilibrium particles.

\section{The kinetic equation for nonequilibrium particles}
\subsection{Solution of the kinetic equation}
The energy-balance equation (\ref{IV.29}) in its turn is determined
by the kinetic equation's solution relative to the non-equilibrium
distribution\\ $\Delta f(t,p)$. Using here he relation (\ref{IV.25})
we can reduce the kinetic equation for superthermal particles
(\ref{Kin_Frid}) with the collision integral (\ref{II.22}) to form:
\begin{equation}\label{IV.31}
\frac{\partial \Delta f_a}{\partial t}= -{\displaystyle
\frac{8\pi N}{3\tilde{p}L(\frac{1}{2}\tilde{p}\ T_0T
N^{1/4})}} {\displaystyle
\left(\frac{2\pi^3}{45}\right)^{1/4}}T^2(t)a(t)\Delta f_a.
\end{equation}
Solving (\ref{IV.31}) we obtain:
\begin{equation}\label{IV.32}
\Delta f_a(t,\tilde{p})=\Delta f^0_a(\tilde{p})
\exp\left[-{\displaystyle\frac{8\pi}{3\tilde{p}}\left(\frac{2\pi^3}{45}\right)^{1/4}
\int\limits_0^t \frac{N aT^2 dt}{L(\frac{1}{2}\tilde{p}\
T_0T N_0^{1/4})}}\right],
\end{equation}
where
\begin{equation}\label{IV.33}
\Delta f^0_a(\tilde{p})\equiv \Delta f_a(0,\tilde{p}).
\end{equation}

\subsection{Transformation to the dimensionless normalized variables}
Let us introduce the {\it average conformal energy of
ultrarelativistic particles' non-equilibrium component at the
initial time},$\langle \tilde{p}\rangle_0$, --
\begin{equation}\label{IV.34}
\langle
\tilde{p}\rangle_0=\frac{\tilde{\varepsilon}(0)}{\tilde{n}(0)}\equiv
{\displaystyle%
\frac{\sum\limits_a (2S+1)\int\limits_0^\infty \Delta
f^0_a(\tilde{p})\tilde{p}^3d\tilde{p}}{\sum\limits_a
(2S+1)\int\limits_0^\infty \Delta
f^0_a(\tilde{p})\tilde{p}^2d\tilde{p}} %
}
\end{equation}
and the {\it dimensionless normalized momentum variable}, $\rho$, --
\begin{equation}\label{IV.35}
\rho\equiv \frac{\tilde{p}}{\langle\tilde{p}\rangle_0},
\end{equation}
so that
\begin{eqnarray}\label{IV.36}
\langle \rho\rangle_0=
\frac{\tilde{\varepsilon}(0)}{\langle
\tilde{p}\rangle_0\tilde{n}(0)}\equiv 1 \Rightarrow \nonumber\\
\langle \rho\rangle_0={\displaystyle%
\frac{\sum\limits_a (2S+1)\int\limits_0^\infty \Delta
f^0_a(\rho)\rho^3d\rho}{\sum\limits_a (2S+1)\int\limits_0^\infty
\Delta f^0_a(\rho)\rho^2d\rho}}=1.
\end{eqnarray}
According to the mathematical model of non-equilibrium plasma, the
average energy of particles in the initial non-equilibrium
distribution must be higher and even much higher then the thermal
energy of particles. Thus, relative to (\ref{IV.25}), (\ref{IV.34})
in the considered model it is:
\begin{equation}\label{IV.37}
\langle \tilde{p}\rangle_0 \gg 1.
\end{equation}
Value $\langle \tilde{p}\rangle_0$ in fact is an {\it independent
parameter} of the considered model and the physical meaning of this
dimensionless parameter is the relation of the average energy of the
initial particles' non-equilibrium distribution to the temperature
of plasma in the equilibrium Universe at the initial time
These values themselves can be infinite but their relation
is finite.
As opposed to the conformal momentum variable
$\tilde{p}$ the average value of the dimensionless conformal
momentum variable $\rho$ is identically equal to $1$.

Let us transform an expression in the exponent of (\ref{IV.32}),
processing to the dimensionless variables $y,\rho$. Taking into
account the weak dependency of the logarithmical factor $L$ on its
arguments and the decreasing character of the integrand in
(\ref{IV.32}), we accept the following estimation of the
logarithmical factor:
\begin{equation}\label{IV.39}
L\left(\frac{1}{2}\tilde{p}\ T_0T N_0^{1/4}\right)\simeq
L(\langle\tilde{p}\rangle_0 T_0^2)\equiv L_0(t).
\end{equation}
Thus, we represent the solution (\ref{IV.32}) in compact form with a
logarithmical accuracy:
\begin{equation}\label{IV.40}
\Delta f_a(t,\rho)=\Delta f^0_a(\rho)
\exp\left(-\frac{2}{\rho}\int\limits_0^t \xi\frac{y^2}{a}dt
\right),
\end{equation}
there the denotation is introduced:
\begin{equation}\label{IV.41}
\xi(t)=\left(\frac{5\pi}{18}\right)^{1/4}
\frac{N}{\langle
\tilde{p}\rangle_0 N_0^{1/2}L_0(t)}\approx
\frac{0.967 N}{\langle
\tilde{p}\rangle_0 N_0^{1/2}L_0(t)}
\approx\frac{N}{\langle
\tilde{p}\rangle_0 N_0^{1/2}L_0(t)}.
\end{equation}

Introducing now the new {\it dimensionless time variable}, $\tau$,
--
\begin{equation}\label{IV.42}
\tau=2\int\limits_0^t \frac{\xi}{a}\ dt\ ,
\end{equation}
such that:
\begin{equation}\label{IV.43}
\frac{d\tau}{dt}\equiv 2 \frac{\xi}{a}>0,
\end{equation}
and the new {\it dimensionless function}, $Z(\tau)$, --
\begin{equation}\label{IV.44}
Z(\tau)=\int\limits_0^\tau y^2(\tau)d\tau ,
\end{equation}
we reduce the kinetic equation's solution (\ref{IV.40}) to form:
\begin{equation}\label{IV.45}
\Delta f_a(\tau,\rho)=\Delta f^0_a(\rho) \cdot{\rm
e}^{\displaystyle - \frac{Z(\tau)}{\rho}}.
\end{equation}

We investigate {\it the coupling equation} (\ref {IV.42}) for 
the dimensionless time variable $ \tau $ and cosmological time $t $. 
Assuming the power dependency of the scale factor $a (t) $ 
on the cosmological time in (\ref {IV.42}) and taking into account the weak dependency of $ \xi $ factor on time, we receive:
\begin{equation}\label{IV.46}
a\sim t^\alpha \rightarrow \left\{ %
\begin{array}{ll}
t^{1-\alpha}, & \alpha\not=1,0;\\
\tau\sim\ln t, & \alpha=1.\\
\end{array}\right.
\end{equation}
Hence it follows that at $\alpha\leqslant 1 \to $
$\tau(\infty)=\infty$, and at $\alpha<1\to$
$\tau(\infty)=\tau_\infty<\infty$. Comparing the relation
(\ref{IV.46}) with the relations (\ref{IV.5}) -- (\ref{IV.8}), we
arrive to the next important conclusion:
\begin{equation}\label{IV.47}
\begin{array}{ll}
\varkappa\geqslant -\frac{1}{3} & (\Omega\leqslant0) \Rightarrow
\tau(\infty)=+\infty;\\[10pt] %
\varkappa<-\frac{1}{3} & (\Omega>0)  \Rightarrow
\tau(\infty)=\tau_\infty<+\infty.%
\end{array}
\end{equation}
Since the distribution function of cosmological plasma's
non-equilibrium component(\ref{IV.45}) depends only on time via the
{\it monotonously increasing function} of variable $Z(\tau)$, the
relations (\ref{IV.47}) mean that in ultrarelativistic cosmological
plasma in the Universe with negative acceleration the total
thermodynamic equilibrium is attained asymptotically, while {\it in
the accelerating Universe it is never attained strictly}.

\subsection{Conformal energy density of the non-equilibrium component}
Substituting the solution of the kinetic equation in form of
(\ref{IV.45}) into expression for the conforaml energy density of
non-equilibrium particles, we obtain
\begin{equation}\label{IV.49}
\tilde{\varepsilon}_{ne}=\frac{45}{64\pi^5}\sum\limits_a (2S+1)
\int\limits_0^\infty \tilde{p}^3\Delta f^0_a(\rho) {\rm e}^{ -
\frac{Z(\tau)}{\rho}}.
\end{equation}
Let us carry out an identical transformation of the given
expression, taking into account the fact that according to
definition (\ref{IV.28}) and the energy-balance equation
(\ref{IV.29}):
\begin{equation}\label{IV.50}
\tilde\varepsilon_{ne}^0=(1-\sigma_0)\frac{3}{32\pi}:
\end{equation}
\begin{equation}\label{IV.51}
\tilde\varepsilon_{ne}\equiv
\frac{\tilde\varepsilon_{ne}}{\tilde\varepsilon_{ne}^0}\tilde\varepsilon_{ne}^0=(1-\sigma_0)\Phi(Z)\frac{3}{32\pi},
\end{equation}
where subject to the transformation to the dimensionless momentum
variable $\rho$ (\ref{IV.35}) we introduce the new {\it
dimensionless function} $\Phi(Z)$:
\begin{equation}\label{IV.52}
\Phi(Z)\equiv {\displaystyle \frac{\sum\limits_a (2S+1)
\int\limits_0^\infty d\rho\rho^3\Delta f^0_a(\rho) {\rm
e}^{\displaystyle - \frac{Z(\tau)}{\rho}}}%
{\sum\limits_a (2S+1) \int\limits_0^\infty d\rho\rho^3\Delta
f^0_a(\rho) }}.
\end{equation}

\subsection{Solution and analysis of the energy-balance equation}

As a result of definition  (\ref{IV.44}) function $Z(\tau)$
satisfies following conditions:
\begin{eqnarray}\label{IV.53}
Z'(\tau)=y^2(\tau)\Rightarrow Z'\ ^2=\sigma(\tau);\\
\label{IV.54}%
Z(0)=0;\quad Z'(0)=y^2(0)=\sqrt{\sigma_0},
\end{eqnarray}
where
\begin{equation}\label{IV.55}
Z'\equiv \frac{dZ}{d\tau}>0.
\end{equation}

Thus, taking into account (\ref{IV.51}) -- (\ref{IV.53}) the
energy-balance equation (\ref{IV.29})can be rewritten in form of the
differen\-tial equation relative to function $Z(\tau)$:
\begin{equation}\label{IV.56}
y^2+(1-\sigma_0)\Phi(Z) =1\Rightarrow
Z'^2+(1-\sigma_0)\Phi(Z) =1,
\end{equation}
solving which subject to relations (\ref{IV.54}) -- (\ref{IV.55}),
we find a formal solution in the implicit form:
\begin{equation}\label{IV.57}
\int\limits_0^Z \frac{du}{\sqrt{1-(1-\sigma_0)\Phi(u)}}=\tau.
\end{equation}

According to definition(\ref{IV.52}) function $\Phi(Z)$ s a
nonnegative one:
\begin{equation}\label{IV.58}
\Phi(Z)>0, \quad (Z\in [0,+\infty)),
\end{equation}
and
\begin{equation}\label{IV.59}
\Phi(0)=1;\quad \lim_{Z\to+\infty}\Phi(Z)=0.
\end{equation}
Calculating the first and the second derivatives of function
$\Phi(Z)$ by $Z$ and differentiating the relation (\ref{IV.52}) by
$Z$, we find:
\begin{eqnarray}\label{IV.60}
\Phi(Z)'_Z<0, \quad (Z\in [0,+\infty));\\
\label{IV.61}%
\Phi(Z)''>0,\quad (Z\in [0,+\infty)).
\end{eqnarray}

In consequence of (\ref{IV.60}) function $\Phi(Z)$ is strictly
monotonously decreasing one but then as a result of the relations
(\ref{IV.59}) it is limited on the interval:
\begin{equation}\label{IV.62}
\Phi(Z)\in [0,1];\quad (Z\in [0,+\infty)),
\end{equation}
and function's $\Phi(Z)$ graph is concave. As a result of these
properties of function $\Phi(Z)$ equation $\Phi(Z)=\Phi_0$ within
the limit being researched, always has a single and the only one
solution $Z=Z_0$, i,e., {\it mapping $Y=\Phi(Z)$ on the set of
nonnegative numbers is bijective}.

Next, from (\ref{IV.53}) it follows that function $Z(\tau)$
monotonously increases on the interval $\tau\in[0,\tau_\infty]$.
Differentiating relation (\ref{IV.56}) by $\tau$ as a composite
function, we find:
\begin{equation}\label{IV.63}
Z'[2Z''+(1-\sigma_0)\Phi'_Z]=0.
\end{equation}
Hence as a result of  $Z'$ positivity (\ref{IV.56}) let us find the
second derivative:
\begin{equation}\label{IV.64}
Z''=-\frac{1}{2}(1-\sigma_0)\Phi'_Z.
\end{equation}
Therefore in consequence of (\ref{IV.60}) and (\ref{IV.27}) --
(\ref{IV.28}) we obtain from (\ref{IV.64}):
\begin{equation}\label{IV.65}
Z''>0,
\end{equation}
i.e function's $Z(\tau)$ graph is also concave. Then,
differentiating (IV.53), subject to (\ref{IV.65}) we find:
\begin{equation}\label{IV.66}
y'>0,
\end{equation}
--- i.e. function $y(\tau)$ (and function $\sigma(\tau)$ together with it)
is a monotonously increasing one. From the other hand it is limited
from below by the initial value $y_0$ ($\sigma_0$), and is limited
from the above by value $1$:
\begin{equation}\label{IV.66}
y'>0,y\in [y_0,1);\quad \sigma'>0, \sigma\in [\sigma_0,1).
\end{equation}

Mentioned properties of functions $y(\tau)$, $Z(\tau)$ and $\Phi(Z)$
assert the bijectivity of chain of mappings $\tau \leftrightarrow
y$, $y \leftrightarrow Z$, $Z \leftrightarrow \Phi$. Finally, each
value $\Phi$ has a one and only one corresponding value $Z$ and one
and only one value $\tau$: $\tau\leftrightarrow \Phi$. To close this
chain it is enough to determine functions $y(\tau)$ and $Z(\tau)$
coupling by means of the energy-balance equation (\ref{IV.57}):
\begin{equation}\label{IV.67}
y=[1-(1-\sigma_0)\Phi(Z)]^{1/4}.
\end{equation}

Equations (\ref{IV.57}) and (\ref{IV.67})  are the {\it parametric
solution of the energy-balance equation} (\ref{IV.56}), and above
mentioned properties of functions $\Phi(Z)$ and $Z(\tau)$ assert the
{\it uniqueness of the solution}. Accor\-ding to (\ref{IV.52})
function $\Phi(Z)$ is fully determined by the initial distribution
of nonequilibrium particles $\Delta f^0_a(\rho)$. Therefore from the
mathematical point of view the problem of thermodynamical
equilibrium recreation in Universe with arbitrary acceleration is
completely solved. Concrete models are determined by dark matter
model and model of initial nonequi\-lib\-rium distribution of
particles.

\begin{figure}[h]
 \centerline{\includegraphics[width=6.8cm]{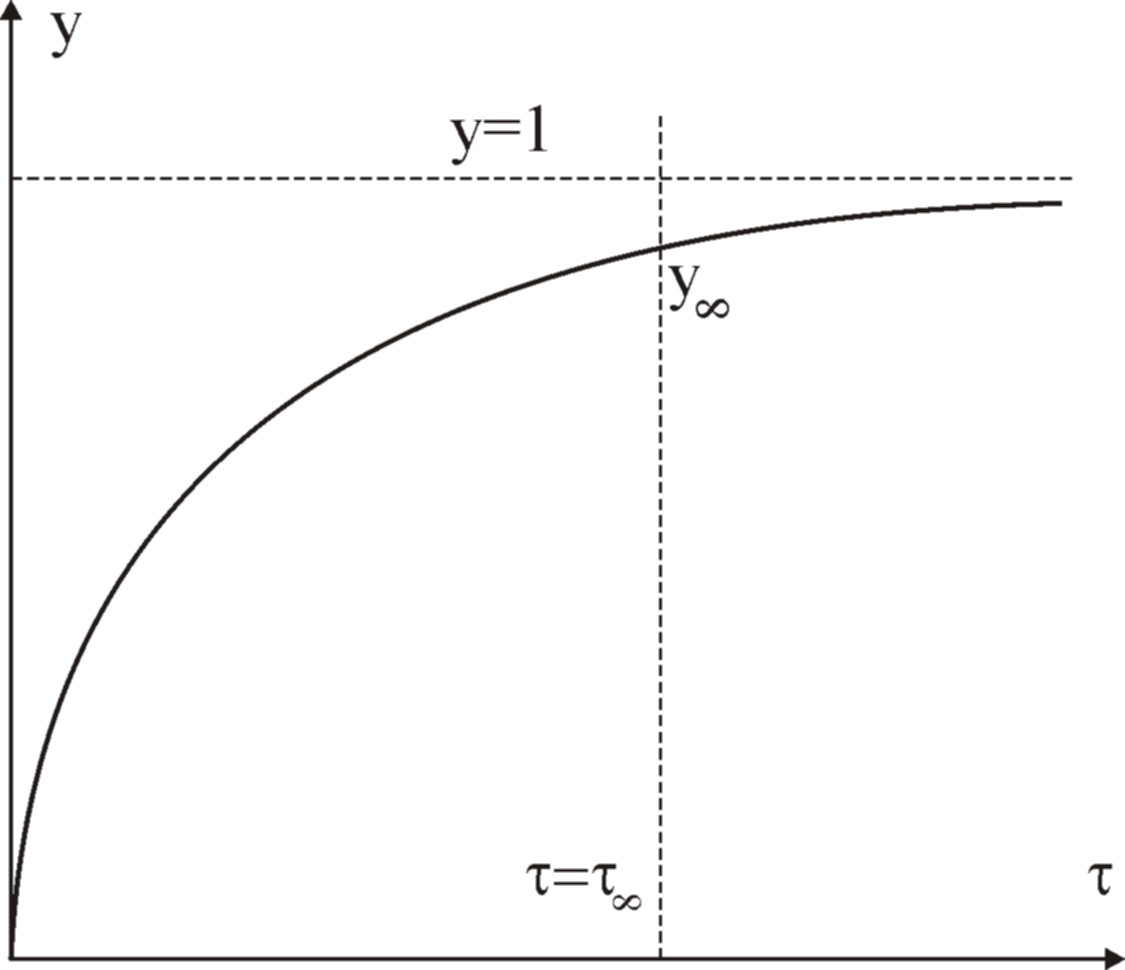}} \caption{Qualitative form of function's
$y(\tau)$ graph.
 \label{graphic_Fig}}
\end{figure}

Let us differentiate now the relation (\ref{IV.64}) by $\tau$ and
take account of the connection (\ref{IV.53}) between functions
$y(\tau)$ and $Z(\tau)$:
\begin{equation}
Z'''=-\frac{1}{2}(1-\sigma_0)\Phi''_{ZZ}Z'
\Rightarrow
y''y=-y'^2-\frac{1}{4}(1-\sigma_0)\Phi''_{ZZ}y^2.
\end{equation}
Thus, as a result of(\ref{IV.61}):
\begin{equation}\label{IV.68}
y''<0,
\end{equation}
--- i.e. graph of function $y(\tau)$, and
$\sigma(\tau)$ graph together with it, are concave. Then since
$\Phi_Z(Z\to\infty)=0$, from (\ref{IV.64}) it follows:
\begin{equation}\label{IV.69}
\lim_{\tau\to\infty}y'(\tau)=0 \Rightarrow
\lim_{\tau\to\infty}\sigma'(\tau)=0,
\end{equation}
--- i.e., value $\sigma=1$ is attained asymptotically at
$\tau\to\infty$. This allows to draw a qualitative graph of
functions $y(\tau)$ (Figure \ref{graphic_Fig}). The finiteness of
the dimensionless time $\tau_\infty$ conducts to the establishment
of the limiting value of function $y(\tau)$:
\begin{equation}\label{IV.70}
y(\tau_\infty)=y_\infty <1 \Rightarrow
\lim_{t\to\infty}y(t)=y_\infty<1.
\end{equation}
In consequence of that the certain part of comsolo\-gi\-cal plasma
energy is forever conserved in the nonequi\-lib\-rium superthermal
component:
\begin{equation}\label{IV.71}
\lim_{t\to\infty}\frac{\varepsilon_{ne}(t)}{\varepsilon_p(t)}=1-\sigma_\infty
= \left\{\begin{array}{ll}
=0, & \tau_\infty=\infty\\
>0, &  \tau_\infty<\infty
\end{array}\right.
.
\end{equation}
According to (\ref{IV.47}) this is possible only for the accelerated
expanding Universe.

\section{Exact model of transition from the ultrarelativistic stage to the inflationary one}
Let us consider a simple model of matter consisting of 2 components (see for details \cite{Ex_Yu,Ign_Arx,Yu_GC13}) -- minimally coupled massive scalar field (cosmological member) with
a state equation:
\begin{equation}\label{V.13}
p_s=-\varepsilon_s,
\end{equation}
and the ultrarelativistic plasma with the state equation
(\ref{IV.12}). Then summary barotropic factor and invariant
acceleration can be written in form:
\begin{equation}\label{V.13}
\varkappa(t)=\frac{1}{3}\frac{1-3\delta}{1+\delta};\quad
\Omega(t)=-\frac{1-\delta}{1+\delta},
\end{equation}
where
\begin{equation}\label{V.13}
\delta=\delta(t)=\frac{\varepsilon_s}{\varepsilon_p} .
\end{equation}

Thus at $\delta={\rm Const}$ formulas(\ref{IV.7}) can be written in
the following convenient form \cite{Ign_Arx,Yu_GC13}:
\begin{equation}\label{V.14}
a=a_1 t^{(1+\delta)/2};\; \varepsilon=\frac{3}{32\pi}
\frac{(1+\delta)^2}{t^2}, \quad \Omega<1.
\end{equation}

\begin{figure}[h]
 \centerline{\includegraphics[width=8cm,height=6cm]{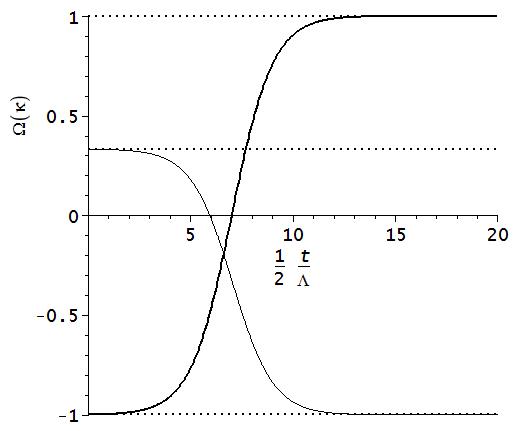}} \caption{Evolution of the effective barotropic factor
$\kappa(t)$ (thin line) and invariant acceleration $\Omega(t)$
(heavy line) relative to the exact solution (\ref{V.21}) at
$\Lambda=1$.  Asymptotes $-1; \; 1/3; \;1$ are denoted by the dotted
lines.
 \label{omega-kappa_Fig}}
\end{figure}

Energy conservation laws (\ref{IV.10}) -- (\ref{IV.11}) take form:
\begin{eqnarray}\label{V.30}
\varepsilon_s={\rm Const}=\frac{3\Lambda^2}{8\pi};\\%
\label{V.18}%
\varepsilon_p a^4 \equiv \tilde{\varepsilon}_p={\rm Const}\simeq
\frac{3}{32\pi}.
\end{eqnarray}
Substituting (\ref{V.30})-(\ref{V.18} )into the equation
(\ref{IV.1}) and integra\-ting it, we obtain:
\begin{equation}\label{V.19}
\!\! a(t)=\frac{1}{\sqrt{2}}\left[
\left(t_0+\sqrt{t^2_0+b^2}\right){\rm e}^{(t-t_0)/2\Lambda}- \frac{b^2}{t_0+\sqrt{t^2_0+b^2}}{\rm e}^{-(t-t_0)/2\Lambda}
\right]^{\frac{1}{2}},
\end{equation}
where:
\begin{equation}\label{V.20}
b^2=\frac{3}{32\pi\Lambda^2}.
\end{equation}
Hence we have, in particular, for the scale factor at $t_0=0$:
\begin{equation}\label{V.21}
a(t)=\frac{1}{\Lambda}\sqrt{\frac{3}{32\pi}{\rm sh} \frac{t}{2\Lambda}}
\end{equation}
Calculating according to (\ref{V.13}), (\ref{V.30}), (\ref{V.18})
and (\ref{V.21}) the relation $\delta$, we find:
\begin{equation}\label{V.24}
\delta(t)=\left(\frac{3}{16\pi\Lambda}\ {\rm sh}\frac{
t}{2\Lambda}\right)^2.
\end{equation}
Next, according to(\ref{V.13}) it is possible to calculate an
effective barotropic factor and an invariant accele\-ra\-tion (see
Figure  \ref{omega-kappa_Fig}).

Following figure shows that by means of para\-me\-ter $\Xi$ it is
easy to control the time of transition to the inflationary
acceleration regime $\kappa\to -1$. Let us recall that cosmological
time $t$ is measured in Planck units.

Thus according to(\ref{IV.42}) we can determine the new
dimensionless time variable $\tau$:
\begin{equation}\label{V.31}
\tau=\frac{2\Lambda\langle\xi\rangle}{\langle\tilde{p}\rangle_0}{\rm F}(\varphi,1/\sqrt{2}),
\end{equation}
where:
\begin{equation}\label{V.32}
\varphi=\arccos \frac{1-{\rm sh}\; t/2\Lambda}{1+{\rm sh}\; t/2\Lambda};
\end{equation}
${\rm F}(\varphi,k)$ is an elliptic integral of the first type (see
e.g.:
\begin{equation}\label{V.34}
{\rm F}(\varphi,k)=\int\limits_0^\varphi
\frac{d\alpha}{\sqrt{1-k^2\sin^2\alpha}};\quad (k^2<1).
\end{equation}

Thus:
\begin{equation}\label{V.35}
\frac{d\tau}{dt}=\frac{1}{2\Lambda}\frac{1}{{\rm
sh}\;\varphi}\frac{{\rm ch}\;\varphi }{1+{\rm sh}\;\varphi}>0;
\quad  \tau\in [0,\tau_\infty),
\end{equation}
where
\begin{equation}\label{V.36}
\tau_\infty=\lim\limits_{t\to+\infty}\tau(t)=\frac{2\Lambda\langle\xi\rangle}{\langle\tilde{p}\rangle_0}
{\rm F}(1,1/\sqrt{2}),
\end{equation}
$F(1,1/\sqrt{2}\approx 1.083216773$.

\begin{figure}[h]
 \centerline{\includegraphics[width=8cm,height=6cm]{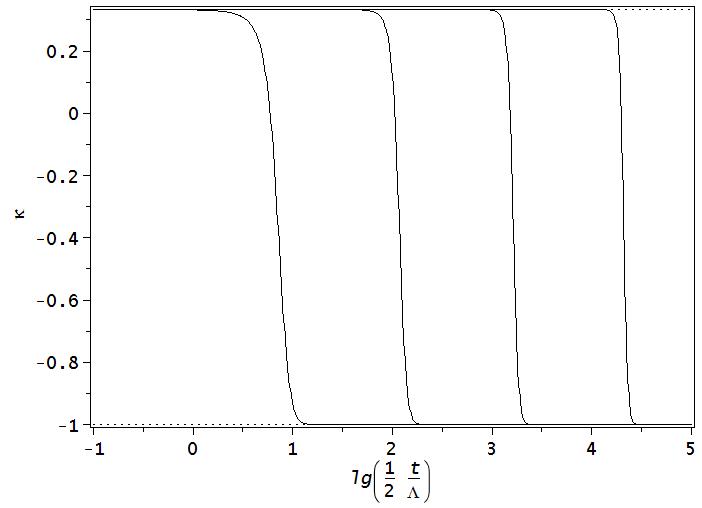}} \caption{The evolution of the barotrope effective coefficient
$\kappa(t)$ relative to the exact solution (\ref{V.21}) subject to
the cosmological constant $\Lambda$ left to right: $\Lambda=1$,
$\Lambda=10$, $\Lambda=100$, $\Lambda=1000$.
 \label{kappass_Fig}}
\end{figure}

\section{The numerical model of LTE restoration in the accelerating Universe}
\subsection{The model of the initial non-equilibrium distribution}
Thus, as it has been mentioned above, the mathema\-tical model of LTE
restoration process in cosmologi\-cal plasma is reduced to two
parametric equations
\begin{equation}\label{IV.57}
\int\limits_0^Z \frac{du}{\sqrt{1-(1-\sigma_0)\Phi(u)}}=\tau.
\end{equation}
\begin{equation}\label{IV.67}
y=[1-(1-\sigma_0)\Phi(Z)]^{1/4},
\end{equation}
which {\it at
given function} $\Phi(Z)$ define the relations of form:
\begin{eqnarray}\label{tau_Z}
\tau=\tau(Z);\\
y=y(Z),
\end{eqnarray}
solving which we can determine function $y(\tau)$ and thereby
formally solve stated problem completely. Thus, the final solution
of the task is found in quadratures specifying the initial
distribution of non-equilibrium particles $\Delta f_0(p)$ and
following defi\-ni\-tion of the integral function $\Phi(Z)$:
\begin{equation}\label{IV.52}
\Phi(Z)\equiv {\displaystyle \frac{\sum\limits_a (2S+1)
\int\limits_0^\infty d\rho\rho^3\Delta f^0_a(\rho) {\rm
e}^{\displaystyle - \frac{Z(\tau)}{\rho}}}%
{\sum\limits_a (2S+1) \int\limits_0^\infty d\rho\rho^3\Delta
f^0_a(\rho) }}.
\end{equation}
Let us note that formally parametric equations
(\ref{IV.57}) and (\ref{IV.67}), as well as function's $\Phi(Z)$
definition, do not differ from the similar, obtained earlier by the
Author in articles \cite{Yu_1986}, \cite{LTE2}. The main new
statement is brought by the acceleration of the Universe and
consists in the relation $\tau(t)$ (\ref{IV.42}).

In order to construct a numerical model let us consider the initial
distribution of white noise type:
\begin{equation}\label{df0}
\Delta f_0(\rho)=\frac{A}{\rho^3}\chi(\rho_0-\rho),
\end{equation}
where $A$ is a normalization constan, $\rho_0>1$ is a dimensionless
parameter, $\chi(x)$ is a Heaviside step function, so that the
conformal energy density with respect to this distri\-bu\-tion is equal
to:
\begin{equation}\label{e_0}
\tilde{\varepsilon}^0_{ne}=\frac{\langle
\tilde{p}\rangle_0^4 A\rho_0}{32\pi^5}.
\end{equation}
Calculating function $\Phi(Z)$ relative to distribution (\ref{df0}),
we find:
\begin{equation}\label{Phi_Z}
\Phi(Z)=e^{-x}-x{\rm Ei}(x);\quad x\equiv \frac{Z}{\rho_0},
\end{equation}
where ${\rm Ei}(x)$ is an integral exponential function
\begin{equation}\label{Ei}
{\rm Ei}(x)=\int\limits_{-1}^\infty \frac{e^{-tx}}{t}dt.\nonumber
\end{equation}
\subsection{The results of numerical integration}
Thus, the problem is reduced to the numerical integration of the
system of equations (\ref{IV.42}), (\ref{IV.57}), (\ref{IV.67}).
Below the certain integration results are represented. Further
according to (\ref{IV.10})--(\ref{IV.12})
\begin{eqnarray}\label{V.30}
\varepsilon_s={\rm Const}=\frac{3\Lambda^2}{8\pi};\\%
\label{V.18}%
\varepsilon_p a^4 \equiv \tilde{\varepsilon}_p={\rm Const}\simeq
\frac{3}{32\pi}.
\end{eqnarray}
and (\ref{V.21})
\begin{equation}\label{V.21}
a(t)=\frac{1}{\Lambda}\sqrt{\frac{3}{32\pi}{\rm sh} \frac{t}{2\Lambda}}
\end{equation}
it may be convenient to
introduce a {\it time cosmological constant}
\begin{equation}\label{t_0}
t_0\equiv 4\Lambda.
\end{equation}
In article \cite{Numer} there was described Author's program designed with Maple v15 and intended  for the numerical simulation of the presented above mathematical model of thermodynamic
equilibrium's restoration in the Universe. Model included transition to the acceleration stage.
Below we describe the results of numerical modeling in-detail and carry
out the analysis.

\begin{figure}[h]
 \centerline{\includegraphics[width=8cm]{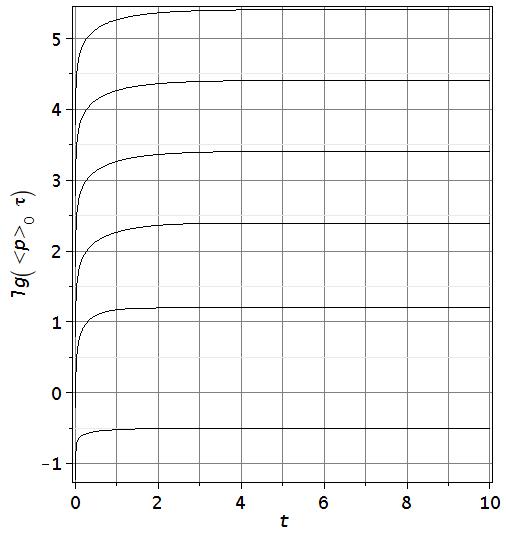}}
 \caption{Dependency of the dimensionless time variable's $\langle\tilde{p}\rangle_0\tau$ common logarithm
 on the cosmological time $t$. Bottom-up: $t_0=1$; $10$; $10^2$; $10^3$; $10^4$; $10^5$. It is introduced everywhere $N_0=100$; $N=10$.
 \label{tau_t_fig}}
\end{figure}

\begin{figure}[h]
 \centerline{\includegraphics[width=8cm]{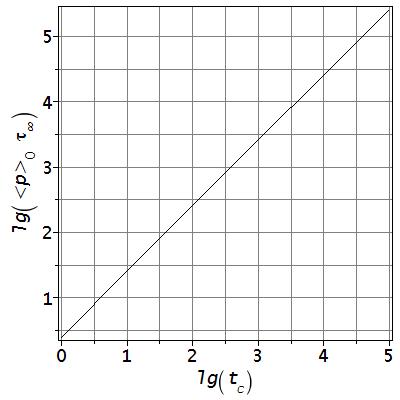}} \caption{Dependency of the dimensionless time parameter $\tau_\infty \langle\tilde{p}\rangle_0$
 on the time cosmological constant $t_0\equiv t_c$ (\ref{t_0}); everywhere is $t_c=1$; $N_0=100$; $N=10$, $\langle\tilde{p}\rangle_0$; $\sigma_0$ --- bottom-up:
 0.01; 0.1; 0.2; 0.5.
 \label{tau8_t0_fig}}
\end{figure}

\begin{figure}[h]
 \centerline{\includegraphics[width=8cm]{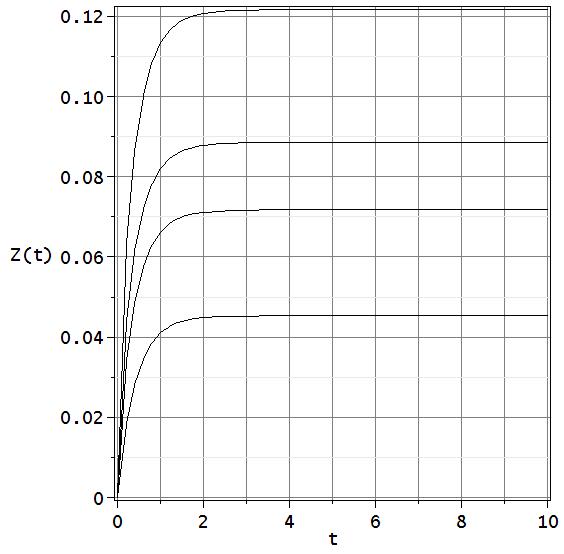}} \caption{Dependency of the dimensionless function $Z(t)$ on time; it is everywhere $t_c=1$; $N_0=100$; $N=10$, $\langle\tilde{p}\rangle_0=10$; $\sigma_0$ --- bottom up:
 0.01; 0.1; 0.2; 0.5.
 \label{Z_t_fig}}
\end{figure}

On Figure  \ref{tau_t_fig} the results of numerical integration for the
definition of the parameter $\tau_\infty$ are shown.
In particular, the integration of the relation (\ref{IV.42}) confirmed
insensitivity of  value $\tau_\infty$ from the number of parameters
and, practically, confirmed the esti\-ma\-tion formula (\ref{IV.42}) \cite{Ex_Yu}
\begin{equation}\label{V.36}
\tau_\infty=\lim\limits_{t\to+\infty}\tau(t)=\frac{2\Lambda\langle\xi\rangle}{\langle\tilde{p}\rangle_0}
{\rm F}(1,1/\sqrt{2}),
\end{equation}
which did not account the details of the logarithmical dependency of
the para\-me\-ter $\langle\xi\rangle$ on time. On Figure  \ref{tau8_t0_fig}
the re\-sults of this value's numerical integration are shown.
These results are well described by formula
\begin{equation}\label{tau_inf}
\tau_\infty\approx \frac{2.57 t_0}{\langle\tilde{p}\rangle_0}.
\end{equation}
On Figure  \ref{Z_t_fig} the dependency of the variable Z(t) at
different values of time cosmological constant is shown.

\begin{figure}
 \centerline{\includegraphics[width=8cm]{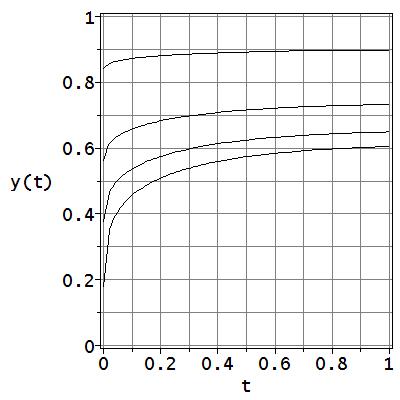}} \caption{Dependency of the relative temperature $y(t)$ on time; it is everywhere $t_c=1$; $N_0=100$; $N=10$, $\langle\tilde{p}\rangle_0=10$; $\sigma_0$ --- bottom up:
 0.01; 0.1; 0.2; 0.5.
 \label{y_t_fig}}
\end{figure}

As it is follows from the results represented on this figure,
function's  $Z(t)$ value also has a limit value at $t\to\infty$:
\begin{equation}\label{Z8}
\lim\limits_{t\to\infty}Z_\infty<\infty.
\end{equation}
According to (\ref{IV.45})
\begin{equation}\label{V.45}
\Delta f_a(\tau,\rho)=\Delta f^0_a(\rho) \cdot{\rm
e}^{\displaystyle - \frac{Z(\tau)}{\rho}},
\end{equation}
this means that at $t\to\infty$
superthermal particles' distribution is ``frozen'':
\begin{equation}\label{IV.45a}
\Delta f_a(\tau_\infty,\rho)=\Delta f^0_a(\rho) \cdot{\rm
e}^{- \frac{Z_\infty}{\rho}}.
\end{equation}
Thus, in modern Universe there can remain the ``tail'' of
non-equilibrium particles of extra-high energies:
\begin{equation}\label{E8}
E\gtrsim E_\infty=Z_\infty\langle\tilde{p}\rangle_0.
\end{equation}

\begin{figure}[h]
 \centerline{\includegraphics[width=8cm]{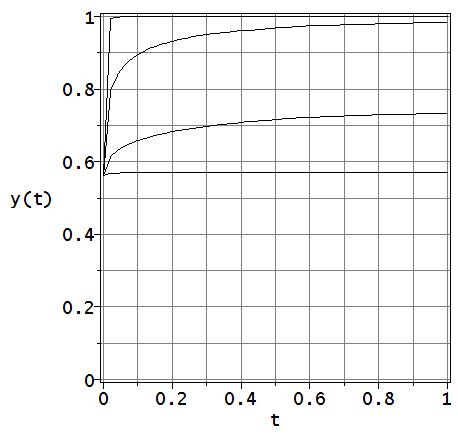}} \caption{Dependency of the relative temperature $y(t)$ on time; it is everywhere $\sigma_0=0.1$; $N_0=100$; $N=10$, $\langle\tilde{p}\rangle_0=10$;  $t_c$ --- bottom-up:
 0.1; 1; 10; 100.
 \label{y_s_fig}}
\end{figure}

\begin{figure}[h]
 \centerline{\includegraphics[width=8cm]{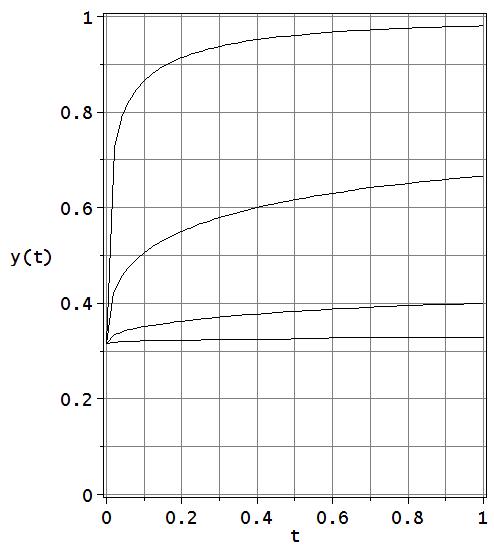}} \caption{Dependency of the relative temperature $y(t)$ on time; it is everywhere  $\sigma_0=0.01$; $N_0=100$; $N=10$, $\langle\tilde{p}\rangle_0=1000$; $t_0$ --- bottom-up:
 1; 10; 100; 1000.
 \label{y_t0_fig}}
\end{figure}

On Figure  \ref{y_s_fig}--\ref{y_t0_fig} the results of numerical
integration for the relative temperature $y(t)=T(t)/T_0(t)\leq1$ are
shown. According to the meaning of that value the dimensionless
parameter:
\begin{equation}\label{E_n8}
e_\infty=1-\sigma_\infty=1-y^4_\infty >0
\end{equation}
is a relative part of cosmological plasma's energy concluded in this
non-equilibrium ``tail'' of distri\-bu\-tion.

\subsection{Asymptotic values of parametres of nonequilibrium distribution at an inflationary stage}
\begin{figure}[h]
 \centerline{\includegraphics[width=8cm]{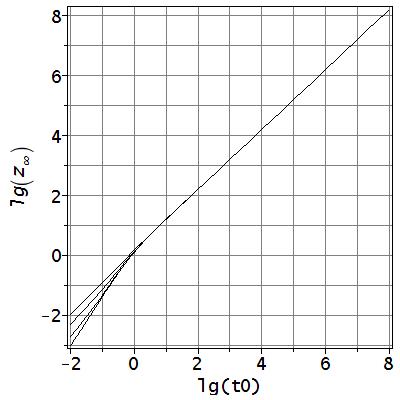}} \caption{Dependency of the dimensionless function $Z_\infty$ on cosmological constant $t_0$; it is everywhere $\langle\tilde{p}\rangle_0=100$; $N_0=100$; $N=10$;  $\sigma_0$ --- bottom-up:
 0.001; 0.01; 0.1; 0.5.
 \label{Z8_t0_fig}}
\end{figure}

\begin{figure}[h]
 \centerline{\includegraphics[width=8cm]{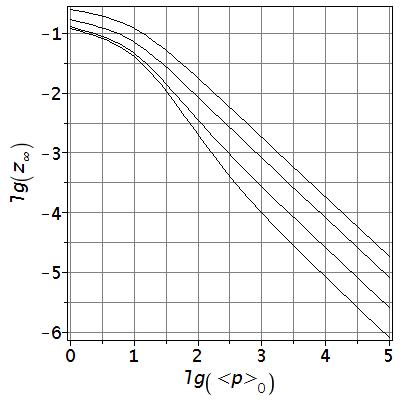}} \caption{Dependency of the dimensionless function $Z_\infty$ on parameter $\langle\tilde{p}\rangle_0$; it is everywhere $t_0=1$; $N_0=100$; $N=10$;  $\sigma_0$ --- bottom-up:
 0.001; 0.01; 0.1; 0.5.
 \label{Z8_lg_p0_fig}}
\end{figure}

\begin{figure}[h]
 \centerline{\includegraphics[width=8cm]{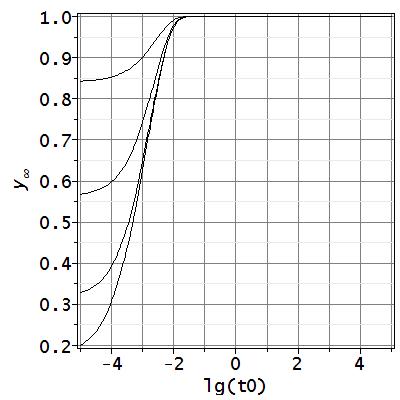}} \caption{Dependency of the relative temperature $y_\infty$ on cosmological constant $t_0$; it is everywhere $t_0=1$; $N_0=100$; $N=10$;  $\sigma_0$ --- bottom-up:
 0.001; 0.01; 0.1; 0.5.
 \label{y8_t0_fig}}
\end{figure}

\begin{figure}[h]
 \centerline{\includegraphics[width=8cm]{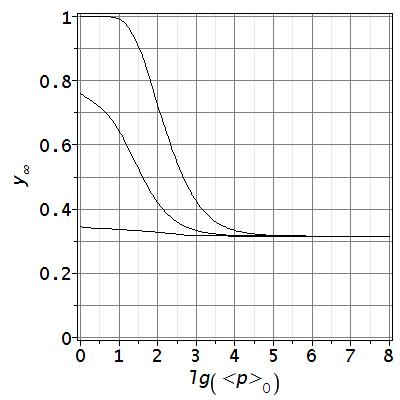}} \caption{of the relative temperature $y_\infty$ on parameter $\langle\tilde{p}\rangle_0$; it is everywhere $\sigma_0=0.1$; $N_0=100$; $N=10$;  $t_0$ --- bottom-up:
 0.1; 1; 10.
 \label{y8_p0_fig}}
\end{figure}

\begin{figure}[h]
 \centerline{\includegraphics[width=8cm]{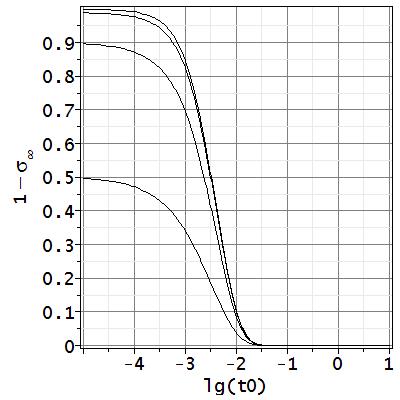}} \caption{Dependence of a relative part of energy of nonequilibrium particles, $1-\sigma_\infty$, on cosmological constant $t_0$; it is everywhere  $\langle\tilde{p}\rangle_0=1000$; $N_0=100$; $N=10$,  $\sigma_0$ --- top-down: 0.001; 0.01; 0.1; 0.5.
 \label{1-s8_t0_fig}}
\end{figure}

\begin{figure}[h]
 \centerline{\includegraphics[width=8cm]{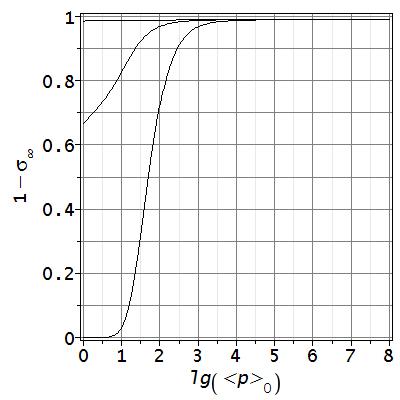}} \caption{Dependence of a relative part of energy of nonequilibrium particles, $1-\sigma_\infty$, on cosmological constant $t_0$; it is everywhere  $\sigma_0=0.01$; $N_0=100$; $N=10$, $\langle\tilde{p}\rangle_0=1000$; $t_0$ --- top-down: 0.1; 1; 10.
 \label{1-s8_p0_fig}}
\end{figure}

For the observation in present age of the Universe the knowledge of possible limiting values of parametres of nonequilibrium distribution of particles is important. Such probable observable parametres are the relative temperature $y_\infty$, the relative part of energy concluded in the nonequilibrium tail of distribution, and also the form of this distribution. On Figure   \ref{Z8_t0_fig}, \ref{Z8_lg_p0_fig}, \ref{y8_t0_fig}, \ref{y8_p0_fig}, \ref{1-s8_t0_fig}, \ref{1-s8_p0_fig} the calculated values of the first two parametres are presented.


\subsection{Analysis of numerical simulation}
From these results it follows, that starting from values of parameter $\langle\tilde{p}\rangle_0$  of the order of 10-100, the survival of considerable number of nonequilibrium relic particles at  modern stage of Universe evolution is possible. It is the striking fact as we recall that according to the results of the Author's early papers, considering standard cosmological scenario which excludes inflationary stage, at the modern stage of expansion it is only relic particles with energy of an order $10^{12}$Gev and above which can survive. At presence of a modern inflationary stage nonequilibrium relic particles with energy of an order of 1 Kev can survive!

On Figure  \ref{de_dr_t_0001_100} the evolution of distribution of density of energy of nonequilibrium particles in the assumption of their initial distribution in form
\begin{equation}\label{de0_1}
\left(\frac{d\varepsilon}{d\rho}\right)_0= A\frac{\chi(\rho_0-\rho)}{(1+\rho/\langle\tilde{p}\rangle_0)^3}
\end{equation}
is represented.

\begin{figure}[h]
 \centerline{\includegraphics[width=8cm]{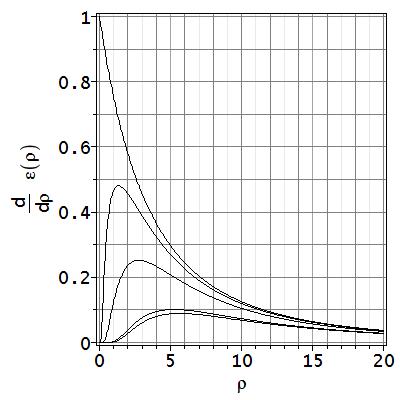}} \caption{Evolution of density of distribution of energy; it is everywhere  $\sigma_0=0.5$; $N_0=100$; $N=10$, $\langle\tilde{p}\rangle_0=10$, $t_0=1$; from top to down:
t= 0.0001; 0.01; 0.1; 1; 10; 100. Lines at t=10 and t=100 are coincide.
 \label{de_dr_t_0001_100}}
\end{figure}

\begin{figure}[h]
 \centerline{\includegraphics[width=8cm]{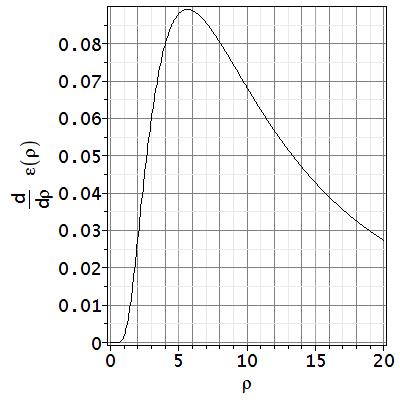}} \caption{Fixing of nonequilibrium function of distribution at $t=\infty$.  $\sigma_0=0.5$; $N_0=100$; $N=10$, $\langle\tilde{p}\rangle_0=10$, $t_0=1$.
 \label{de_dr_t_0001_100}}
\end{figure}

\begin{figure}[h]
 \centerline{\includegraphics[width=8cm]{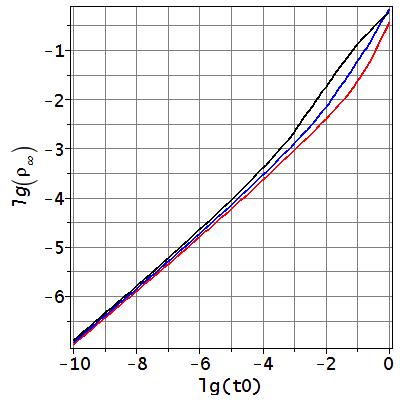}} \caption{Dependence of relativity limiting energy of a maximum $\lg\rho_\infty$ of distribution on a cosmological constant at $\sigma_0=0.01$,  $N_0=100$; $N=10$. Red line --  $\langle\tilde{p}\rangle_0=1$, blue line --  $\langle\tilde{p}\rangle_0=10$, black line --  $\langle\tilde{p}\rangle_0=100$.
 \label{rho8_t0}}
\end{figure}

Calculating a maximum of distribution of density of energy from a relationship (\ref{IV.45}), we will find:
\begin{equation}\label{e_max}
\rho_\infty^{max}=\sqrt{Z_\infty \langle\tilde{p}\rangle_0}.
\end{equation}
On Figure  \ref{rho8_t0} the dependency of the maximum of energy spectrum of nonequilibrium particles on cosmological constant $t_0$ is presented. From this figure it is apparent, that at not very great values of the cosmological constant, the maximum of  energy spectrum of nonequilibrium particles is in the field of quite low energies, that, of course, means their detection is possible in space conditions.
Let's notice, that value of dimensionless energy $\rho$ actually means, that in present period the particle has energy which $\rho\langle\tilde{p}\rangle_0$ times bigger than temperatures of relic radiation.
Thus, in present period we can observe truly relic particles with energies of order of 1 Kev and more in a maximum of the distribution. Speaking about "truly relic particles", we mean the particles which have survived from the moment of the Universe birth, unlike relic photons or neutrino, which were formed at the radiationally-dominated stage of the Universe. Detection of truly relic particles in cosmic space would allow to receive the information about the moment of the Universe birth, and also about  elementary particles interaction specifics at extra-high energies which never would be attainable by mankind.

In the next article we will consider the process of restoration of thermodynamic equilibrium in the Universe with real parameters of the cosmological constant $\Lambda$.
\section*{Acknowledgments}
In conclusion, Author expresses his thanks to professor Vitaly
Melnikov, who had triggered Author's interest for the problem.
Also the Author is grateful to professor Alexey Starobinsky for useful
discussion of problems of cosmological models with acceleration.

\end{document}